\documentclass[aps,pra,twocolumn,amsmath,superscriptaddress,longbibliography]{revtex4-2}

\newcommand{\bea}{\begin{eqnarray}}
\newcommand{\eea}{\end{eqnarray}}
\newcommand{\beq}{\begin{equation}}
\newcommand{\eeq}{\end{equation}}

\usepackage{amsmath}
\usepackage[urlcolor=blue,colorlinks=true,citecolor=blue,linkcolor=blue,pdfstartview={FitH},bookmarks=false]{hyperref}
\usepackage{appendix}
\usepackage{graphicx}
\usepackage{longtable}
\usepackage{epsfig}
\usepackage{dcolumn}
\usepackage{bm}
\usepackage{bbm}
\usepackage{amssymb}
\usepackage{multirow}
\usepackage{times,color}
\usepackage{hyperref}
\usepackage{amsmath}
\usepackage{color}
\usepackage{subfigure}
\usepackage{float}
\usepackage{epstopdf}

\begin{document}
\title{Yang-Lee edge singularity and quantum criticality in non-Hermitian PXP model}

\author{Wen-Yi Zhang}
\affiliation{College of Physics, Nanjing University of Aeronautics and Astronautics, Nanjing, 211106, China}
\affiliation{Key Laboratory of Aerospace Information Materials and Physics (NUAA), MIIT, Nanjing 211106, China}

\author{Meng-Yun Mao}
\affiliation{College of Physics, Nanjing University of Aeronautics and Astronautics, Nanjing, 211106, China}
\affiliation{Key Laboratory of Aerospace Information Materials and Physics (NUAA), MIIT, Nanjing 211106, China}

\author{Qing-Min Hu}
\affiliation{College of Physics, Nanjing University of Aeronautics and Astronautics, Nanjing, 211106, China}
\affiliation{Key Laboratory of Aerospace Information Materials and Physics (NUAA), MIIT, Nanjing 211106, China}

\author{Xinzhi Zhao}
\affiliation{College of Physics, Nanjing University of Aeronautics and Astronautics, Nanjing, 211106, China}
\affiliation{Key Laboratory of Aerospace Information Materials and Physics (NUAA), MIIT, Nanjing 211106, China}
\affiliation{School of Physical Science and Technology, Ningbo University, Ningbo 315211, China}

\author{Gaoyong Sun}
\email{gysun@nuaa.edu.cn}
\affiliation{College of Physics, Nanjing University of Aeronautics and Astronautics, Nanjing, 211106, China}
\affiliation{Key Laboratory of Aerospace Information Materials and Physics (NUAA), MIIT, Nanjing 211106, China}
          
\author{Wen-Long You}
\email{wlyou@nuaa.edu.cn}
\affiliation{College of Physics, Nanjing University of Aeronautics and Astronautics, Nanjing, 211106, China}
\affiliation{Key Laboratory of Aerospace Information Materials and Physics (NUAA), MIIT, Nanjing 211106, China}

\begin{abstract}
We present a comprehensive theoretical framework for quantum criticality in the non-Hermitian detuned PXP model, and establish the complete phase diagram, which had remained elusive in previous studies.  Starting from a numerically identified phase transition point, we construct an exact second-order phase transition boundary through a similarity transformation in the real-energy regime. By introducing the biorthogonal entanglement entropy and biorthogonal Loschmidt echo, we demonstrate from both equilibrium and nonequilibrium perspectives that this transition belongs to the Ising universality class. Using the correlation function, we further distinguish between confined and deconfined phases within the $\mathcal{PT}$-symmetric region. In the complex-energy regime, we identify both a full $\mathcal{PT}$ transition and a first-excited-state $\mathcal{PT}$ transition, respectively. Moreover, we identify the location of the Yang-Lee edge singularity (YLES) using both the associated-biorthogonal and self-normal Loschmidt echoes, and extract the corresponding critical exponent, which agrees with the predictions of non-unitary conformal field theory. Finally, we propose an experimental scheme to observe the YLES in Rydberg atomic arrays, which offers a promising route to exploring non-Hermitian critical phenomena and singularities in future experimental settings.
\end{abstract}

\date{\today}

\maketitle

\section{Introduction}
\label{intro}

In quantum mechanics, the Hamiltonian of a physical system is typically assumed to be Hermitian, %a Hermitian operator,
ensuring the reality of the energy spectrum, 
unitary time evolution, and real-valued physical observables~\cite{v.Neumann1930}. 
However, many realistic systems are inevitably coupled to external environments, leading to particle loss, energy dissipation, {and} measurement backaction, all of which violate conservation laws~\cite{El-Ganainy2018}. 
To describe the dynamics of such open quantum systems, %one commonly employs 
master equation approaches, particularly the Lindblad formalism, are commonly used~\cite{Lindblad1976}. 
By tracing out environmental degrees of freedom, the system’s no-jump evolution is effectively governed by a non-Hermitian Hamiltonian with anti-Hermitian components~\cite{Gorini1976}. 

While non-Hermitian Hamiltonians break unitarity and time-reversal symmetry, they can still exhibit well-defined physical behavior under specific symmetry conditions. 
In particular, when a non-Hermitian Hamiltonian respects parity-time ($\mathcal{PT}$) symmetry, its spectrum can remain entirely real over a finite parameter range~\cite{Bender1999}. 
{As the system parameters are varied, $ \mathcal{PT}$ symmetry may break, leading to a transition from real to complex energy eigenvalues, known as the $ \mathcal{PT}$ transition~\cite{Ashida2020,Kawabata2022}. In non-Hermitian systems, $ \mathcal{PT}$-symmetry breaking behaves differently from the traditional spontaneous symmetry breaking observed in closed quantum systems~\cite{Yamamoto2019}.} This mechanism has been extensively studied experimentally, with realizations in optical waveguide arrays~\cite{Chen2022}, cold atomic systems~\cite{Malossi2014,Kunitski2019}, superconducting circuits~\cite{Koczor2022}, and photonic crystals~\cite{Chen2022PRB}.

Research on non-Hermitian systems has rapidly expanded, revealing a host of phenomena beyond the scope of traditional Hermitian spectral theory, including exceptional points (EPs)~\cite{Hodaei2017,Zhou2018,Xiao2021}, biorthogonal eigenstates~\cite{Chang2020,Sun2021,Sanno2022}, and the non-Hermitian skin effect~\cite{Lee2016,Yao2018,Kunst2018,Borgnia2020}.
These features also motivate a reexamination of the nature of quantum phase transitions.
In non-Hermitian settings, critical behavior is no longer restricted to the closure of the ground-state energy gap~\cite{Matsumoto2020,Yang_2022}. Instead, it may manifest through spectral complexification~\cite{Lu2024PRB}, non-analytic features in excited states~\cite{Zhang2022,Lu2024}, or even dynamical singularities in time-evolved overlap functions~\cite{Tang2022}. 
Among the many unconventional critical phenomena uncovered in non-Hermitian systems, the Yang-Lee edge singularity (YLES) stands out as a prototypical example of non-unitary criticality~\cite{Cardy1985}. 
The YLES, initially introduced in the context of Lee-Yang zeros for classical partition functions under complex fields, represents the critical boundary where these zeros accumulate, signaling a phase transition within the corresponding non-Hermitian framework~\cite{Yang1952,Lee1952}. YLES arises from spectral singularities in the complex plane, often associated with EPs where eigenvalues and eigenvectors coalesce. Its critical behavior falls outside traditional universality classes and is instead governed by non-unitary conformal field theory (CFT)~\cite{Kortman1971,Michael1978}. {The density of Yang-Lee zeros near the edge is a powerful diagnostic of both critical exponents and the order of a phase transition in classical systems~\cite{Dalmazi_2010,Dalmazi2010}, and it has been increasingly applied to selected quantum systems as well~\cite{YinShuai2024}. Recent studies have broadened investigations of the YLES to encompass entanglement-based diagnostics~\cite{Jian2021,Li2025} and superconductors~\cite{Li2023}.} The growing field of non-Hermitian quantum physics has recently enabled experimental access to YLES, revealing their critical behavior through mappings to classical models. Various approaches, including utilizing Kibble-Zurek scaling~\cite{Zhai2018,Dora2019}, exploiting the coherence of a probe spin~\cite{Wei2017}, implementing dissipation in Rydberg atomic arrays~\cite{Shen2023}, using the single-photon interferometric network~\cite{Gao2024}, applying the single-qubit~\cite{Lan2024} and employing the Loschmidt echo~\cite{lu2025} have been proposed to realize these singularities in quantum many-body systems. However, directly observing their critical points and extracting critical exponents in quantum many-body systems remains challenging.

Complementary to the control of spectral structures in non-Hermitian systems, another active research direction focuses on structural constraints imposed on the accessible state space.
In recent years, kinetically constrained dynamics has attracted significant attention in the context of quantum many-body physics.
These models restrict certain local transitions through projection operators, thereby substantially compressing the dimensionality of the accessible Hilbert space and altering the system’s thermalization behavior~\cite{You2022,Zhang2023,Zhang2024,Hu2025}.
A canonical example is the PXP model~\cite{Jaksch2000,Turner2018,Lin2019}, where spin flips are permitted only when both neighboring sites are in the ground state. 
This mechanism has been experimentally confirmed, particularly in Rydberg atom chains, where it emerges naturally from the underlying interaction blockade~\cite{Bernien2017,Ho2019,Giudici2022}.
Furthermore, similar kinetic constraints have been artificially engineered in cold atom platforms via quantum gas microscopes~\cite{Browaeys2020}, and in optical lattices~\cite{Sierant2018,Zhao2020} through controlled quantum jump dynamics.

The emergence of complex energy spectra due to non-Hermitian couplings and the state-space compression introduced by kinetic constraints represent two complementary routes to novel many-body behavior. 
Integrating these mechanisms into a single framework offers an opportunity to explore unconventional quantum phase transitions under experimentally accessible conditions. 
In this work, we propose and investigate a non-Hermitian quantum many-body model that incorporates both kinetic constraints and complex-valued couplings. 
The model incorporates kinetically constrained structures to restrict spin-flip processes, allowing local operations only under specific neighboring configurations; a non-Hermitian coupling term that introduces asymmetric transitions, breaks time-reversal symmetry, and drives spectral complexification; and a uniform detuning term that modulates particle distributions, thereby controlling the structure of the accessible Hilbert space and the energy gap. Each term in the Hamiltonian corresponds to a physically realizable process with a clear implementation scheme in existing experimental platforms.
We systematically investigate the quantum phase transitions of the model and obtain a complete phase diagram that captures the crossover from Hermitian to non-Hermitian regimes. The system exhibits both conventional second-order quantum phase transitions and $ \mathcal{PT}$ transitions unique to non-Hermitian physics. The scaling behavior of the second-order transition is identified and characterized using both equilibrium and nonequilibrium approaches. In addition, by tracking the evolution of the imaginary components of the energy spectrum, we determine both the global $ \mathcal{PT}$ transition and a first-excited-state $ \mathcal{PT}$ transition. Meanwhile, we identify the YLES in the system and extract the critical exponent. This provides a potential pathway toward observing YLESs in experimentally tunable non-Hermitian constrained systems.

This paper is organized as follows. In Sec. \ref{sec:nHmodel}, we introduce
the non-Hermitian detuned PXP model and present its phase diagram. In Sec. \ref{sec:Second-order phase transition}, we discuss the second-order phase transition in the real-energy regime and define the {confined and deconfined} phases. In Sec. \ref{sec:PT transition}, we identify a full $\mathcal{PT}$ transition and a first-excited-state $\mathcal{PT}$ transition in the complex-energy regime, and further locate the YLES, for which we also propose an experimental scheme for observation. Conclusions and discussions are presented in Sec. \ref{sec:Summary and conclusions}.

\section{Non-Hermitian detuned PXP Model and Phase diagram}
\label{sec:nHmodel}

\begin{figure}[tb]
\centering
\includegraphics[width=\columnwidth]{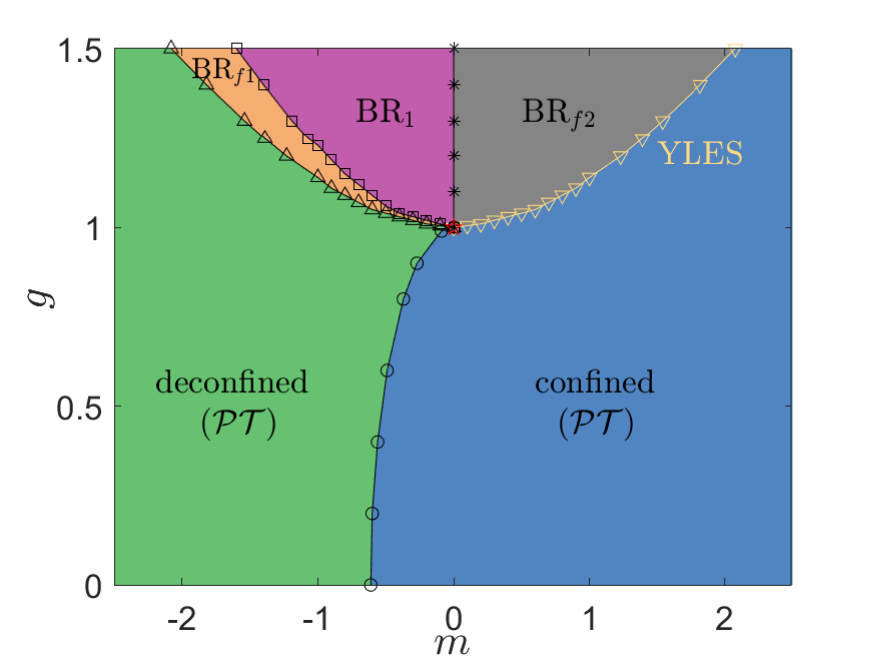}
\caption{{Quantum phase diagram of the non-Hermitian detuned PXP model [Eq.~(\ref{equ:nHdPXP})], corresponding to the $\alpha=\pi/2$ slice of the parent Hamiltonian [Eq.~(\ref{eq:Ising})].   The labels "deconfined" and "confined" refer to the ground-state phases. $\mathcal{PT}$ and BR$_f$ denote  the $\mathcal{PT}$-symmetric and $\mathcal{PT}$-broken regions of the full many-body spectrum, respectively. Two distinct broken regimes are labeled BR$_{f1}$ and BR$_{f2}$. BR$_1$ corresponds to $\mathcal{PT}$ breaking in the first excited state, while the ground state remains $\mathcal{PT}$-symmetric. The red pentagram marks the EP at which all eigenvalues coalesce to zero.  The yellow line indicates the second-order transition line and coincides with the YLES.}
}
\label{fig:phase_diagram1}
\end{figure}

\begin{figure}[tb]
\centering
\includegraphics[width=0.98\columnwidth]{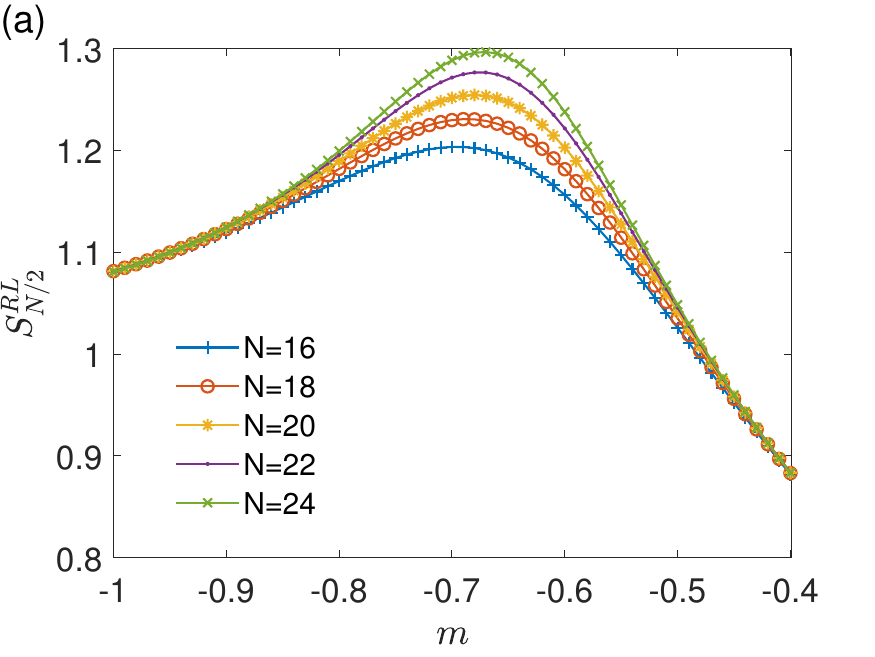}
\includegraphics[width=0.98\columnwidth]{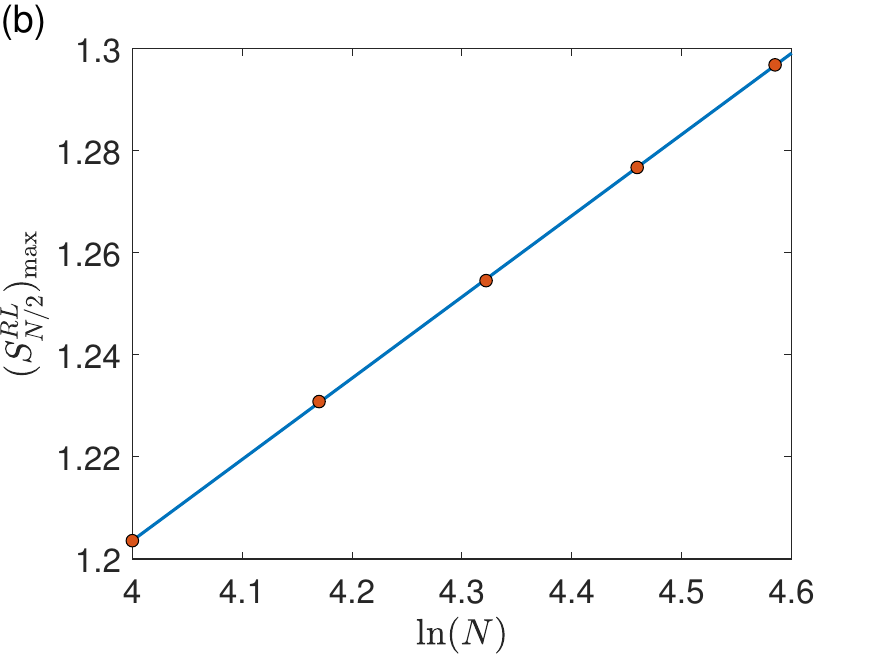}
 \caption{Entanglement entropy for $g=0.1$. (a) The biorthogonal entanglement entropy with respect to $m$ for systems from $N = 16$ to $N = 24$. (b) The finite-size scaling of the biorthogonal entanglement entropy at peaks shown in (a), which are fitted by using Eq. (\ref{eq:central charge}). The central charge is identified as $c = 0.483\pm0.019$ from the biorthogonal entanglement entropy.}
 \label{fig:entropy}
\end{figure} 

To explore the YLES under experimentally accessible conditions, we are motivated by the combined effects of non-Hermitian interactions that induce complex spectra and kinetic constraints that restrict the accessible Hilbert space, {which together give rise} to rich and unconventional many-body behavior. {Accordingly, we consider a representative non-Hermitian Ising chain with transverse and longitudinal fields:
\begin{eqnarray}
\label{eq:Ising}
    \hat{H} \!= \!J \!\sum_{j=1}^{N} \sigma^z_j\sigma^z_{j+1} \!+\!h_x \!\sum_{j=1}^N \sigma^x_j\! +\! ge^{i\alpha} \!\sum_{j=1}^{N} \sigma^y_j\!\!+\!h_z \!\sum_{j=1}^{N} \sigma^z_j,
\end{eqnarray}
where $\sigma^x_j, \sigma^y_j, \sigma^z_j$ denote the Pauli matrices along $x, y, z$ directions at lattice site $j$,} $h_x$ and $h_z$ denote the strengths of the transverse and longitudinal fields respectively, $J$ represents the {strength of ferromagnetic exchange} interaction between neighboring lattice sites, and $g$ characterizes the amplitude of the non-Hermitian term. The parameter $e^{i\alpha}$ allows continuous interpolation between the Hermitian and non-Hermitian regimes. {The non-Hermitian Hamiltonian (\ref{eq:Ising}) with $\alpha=\pi/2$ has $\mathcal{PT}$-symmetry, i.e., $(\mathcal{PT})\hat{H}(\mathcal{PT})^{-1}=\hat{H}$, while $\mathcal{P}\hat{H}\mathcal{P}^{-1}\neq \hat{H}$ and $\mathcal{T}\hat{H}\mathcal{T}^{-1}\neq \hat{H}$ separately. Here the time-reversal operator $\mathcal{T}$ satisfies $\mathcal{T}i\mathcal{T}^{-1}=-i$ and $\mathcal{T}\vec{\sigma}_j\mathcal{T}^{-1}=-\vec{\sigma}_j$. The parity operator $\mathcal{P}$ is defined by rotating each spin by $\pi$ about the $y$-axis, $\mathcal{P}=\prod_{j=1}^N (i\sigma^y_j)$~\cite{Yang_2022}. Since the Hamiltonian is not $\mathcal{PT}$-symmetric for other values of $\alpha$, we henceforth restrict our analysis to $\alpha=\pi/2$. The cases for other values of $\alpha$ are presented in Appendix~\ref{APPENDIX B}.} When $h_z-2J \sim  h_x \ll h_z$, the Hamiltonian, as derived in detail in Appendix~\ref{APPENDIX A}, can be rewritten in a Rydberg-constrained form:
{
\begin{eqnarray}
\label{equ:nHdPXP}
\hat{H} =&& h_x\sum_{j=1}^{N}  P_{j-1} \sigma^x_{j} P_{j+1} +ig \sum_{j=1}^{N}  P_{j-1}\sigma^y_{j} P_{j+1} \nonumber \\ &&+ 2m \sum_{j=1}^N P_{j-1}n_{j}P_{j+1},
\end{eqnarray}
where $2m=h_z-2J$ characterizes the detuning amplitude.} The projection operators $P_j = (1 - \sigma^z_j)/2$ suppress adjacent excitations by enforcing a kinetic constraint similar to the Rydberg blockade, with $n_j = 1-P_j$ denoting the local occupation.  Throughout the main text, we set the transverse field strength $h_x = 1$ unless otherwise specified. This {constrained  model consists of three essential ingredients: kinetic constraints via projection operators $P_j$, which enforce local transition rules based on neighboring configurations; the non-Hermitian modulation term $ig \Sigma_{j=1}^{N}  P_{j-1}\sigma^y_{j} P_{j+1}$, which generates complex dynamics; and a uniform detuning term $\Sigma_{j=1}^N P_{j-1}n_{j}P_{j+1}$, which controls occupation imbalances and the effective energy landscape. {Likewise, the Hamiltonian (\ref{equ:nHdPXP}) is $\mathcal{PT}$-symmetric with the same $\mathcal{P}$ and $\mathcal{T}$, since $P_j$ and $n_j$ are $\mathcal{PT}$-even and the non-Hermitian term appears as $i g \sigma_j^y$.} To analyze Eq.(\ref{equ:nHdPXP}), we introduce a non-unitary} similarity transformation~\cite{Zhang2020,Yang_2022}:
{
\begin{eqnarray}
\label{eq:transformation}
    \tilde{\sigma}^x_j \!\!=\!\! \frac{\sigma^x_j+ig\cdot \sigma^y_j}{\sqrt{1-g^2}},\quad \tilde{\sigma}^y_j \!\!=\!\! \frac{\sigma^y_j-ig\cdot \sigma^x_j}{\sqrt{1-g^2}},\quad \tilde{\sigma}^z_j \!\!=\!\! \sigma^z_j. 
\end{eqnarray}
The {transformed spin operators preserve the  $\mathfrak{su}(2)$ commutation relations}, although $\tilde{\sigma}^x_j$ and $\tilde{\sigma}^y_j$ are not Hermitian.} Applying the transformation to the Hamiltonian (\ref{equ:nHdPXP}), we have
{
\begin{eqnarray}
\label{eq:mathcal_H}
    \tilde{H} \!\!=\!\! \sqrt{1\!\!-\!\!g^2} \! \sum_{j=1}^{N}\!  \tilde{P}_{j-1} \tilde{\sigma}^x_{j} \tilde{P}_{j+1} \!\!+\!\! 2m \! \sum_{j=1}^{N}\!  \tilde{P}_{j-1}\tilde{n}_{j} \tilde{P}_{j+1}.
\end{eqnarray}}

\begin{figure*}[tb]
\centering
\includegraphics[width=0.98\columnwidth]{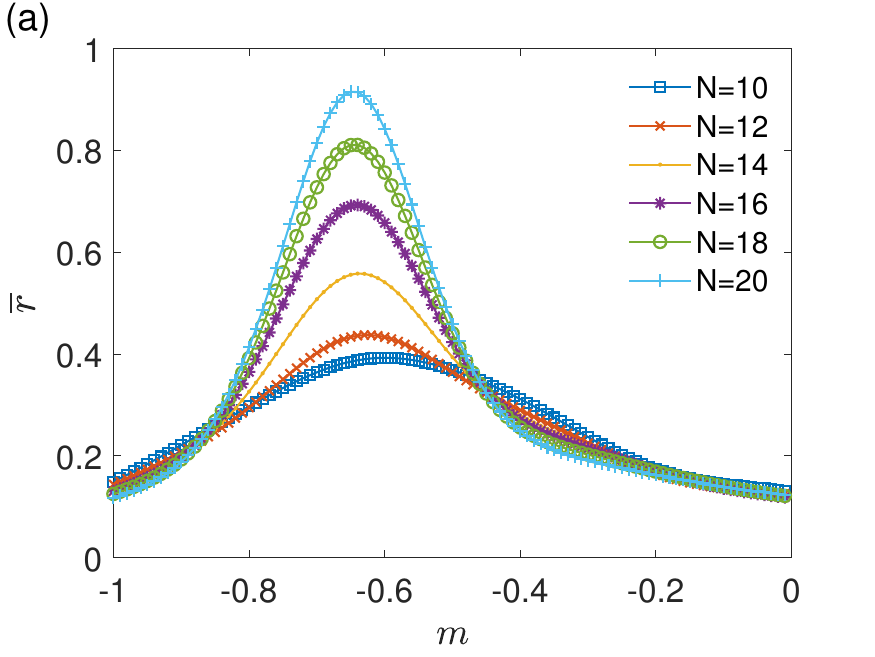}
\includegraphics[width=0.98\columnwidth]{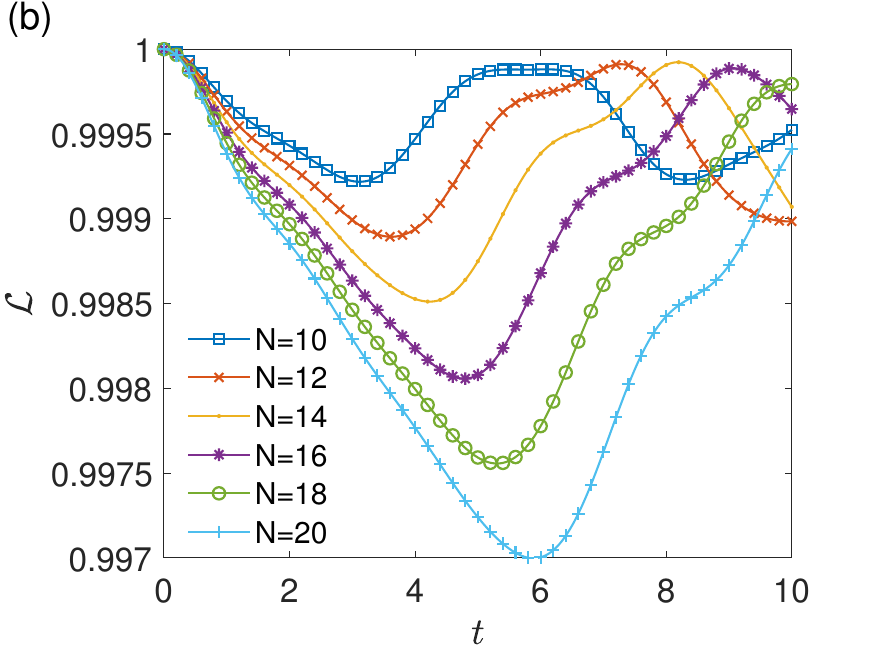}
\includegraphics[width=0.98\columnwidth]{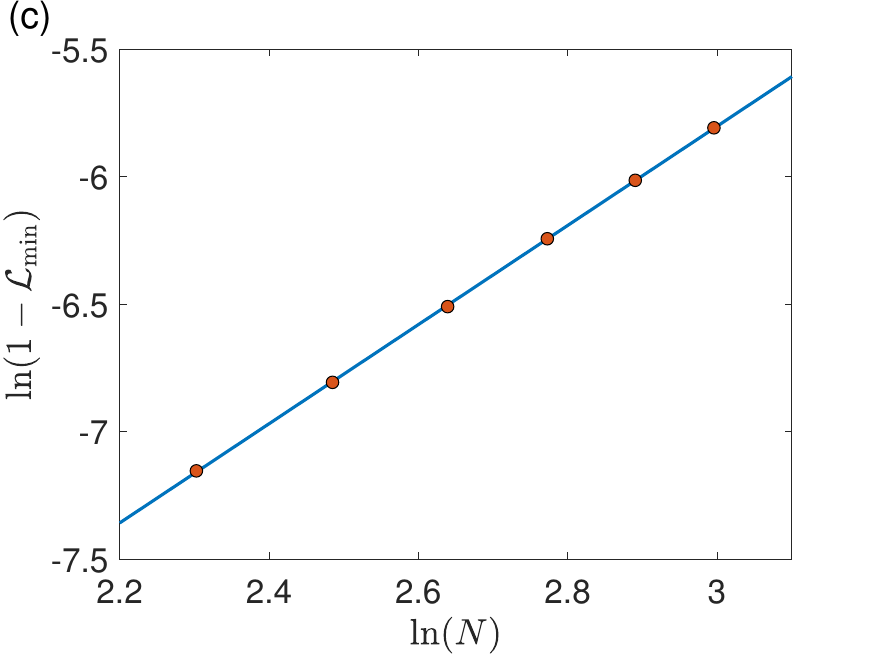}
\includegraphics[width=0.98\columnwidth]{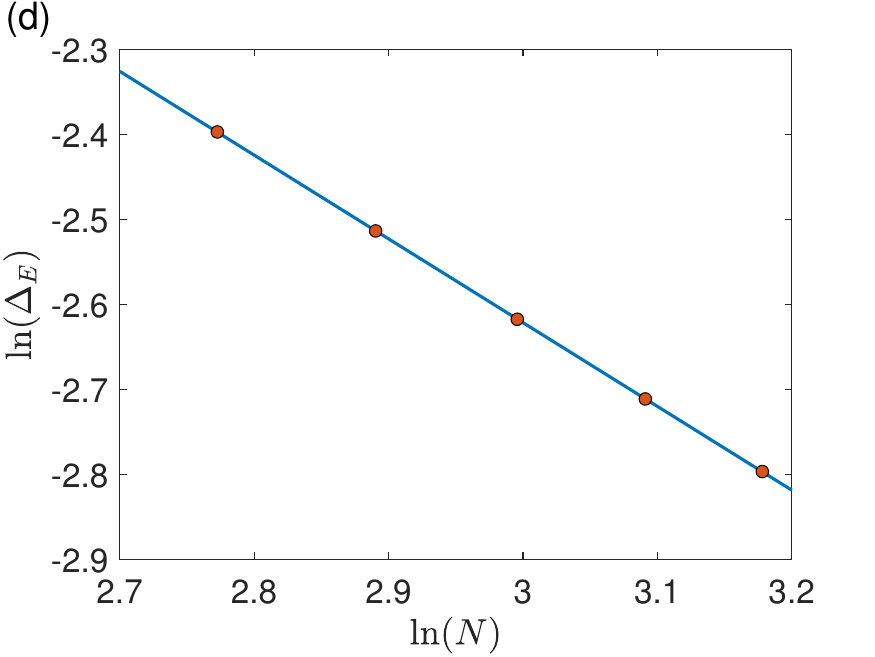}
 \caption{Extraction of the critical exponent. (a) The short-time average rate function for $g=0.1$ with respect to $m$ with $\delta m = 0.01$. (b) The biorthogonal Loschmidt echo at the peak position $m^{(N)}$ of $t$ as a function of time. (c) Finite-size scaling of $(1-\mathcal{L}_{\rm min})$ obtained from panel (b) as a function of lattice sizes $N$. The correlation-length critical exponent $\nu=1.028\pm0.031$ is obtained from the fitting curve. (d) The scaling behavior of $\Delta_E$ versus $N$ around the critical point. The dynamical exponent $z$ obtained from the fitting is $z = 0.986\pm0.034$.}
 \label{fig:Loschmidt echo}
\end{figure*} 
For $g<1$, the Hamiltonian (\ref{eq:mathcal_H}) maintains a fully real energy spectrum. The phase diagram in Fig. \ref{fig:phase_diagram1} includes two ground-state phases, namely the {deconfined and confined} phases. A continuous second-order phase transition separates the two phases as $m$ increases. When $g=0$, the model effectively reduces to the PXP model with an additional uniform detuning. The critical point of this model is known to be $m_c = -0.655$~\cite{Fendley2004,Zhang2023}. Therefore, in the region where $g<1$, the phase boundary in Fig. \ref{fig:phase_diagram1} can be obtained analytically and is given by $m_c=-0.655\sqrt{1-g^2}$. Notably, at the point where $g=1$ and $m=0$, the Hamiltonian vanishes, resulting in all eigenvalues coalescing to zero. This specific point signifies an EP.

For $g>1$, the analytical expression is no longer applicable in this regime. At this stage, in addition to the $\mathcal{PT}$-symmetric phase, $\mathcal{PT}$-symmetry-broken phases also emerge, which we refer to as the BR phases.
In particular, we designate the phase in which the full many-body spectrum breaks $\mathcal{PT}$ symmetry as BR$_f$. When $\mathcal{PT}$ symmetry is broken in the first excited state, while the ground state remains purely real, we refer to this regime as the BR$_1$ phase. Moreover, the BR$_f$ phase can be further divided into two distinct regions. In the BR$_{f1}$ regime, the ground-state energy only has a non-degenerate real part, and the first excited state forms a pair of complex-conjugate eigenvalues. In contrast, the BR$_{f2}$ regime is characterized by a degenerate real part of the ground-state energy, with the corresponding eigenvalues also forming complex-conjugate pairs. 

As the parameter $m$ increases from negative values, the system undergoes a full $\mathcal{PT}$ transition from the $\mathcal{PT}$-symmetric deconfined phase into the BR$_{f1}$ phase. This is followed by a first-excited-state $\mathcal{PT}$ transition into the BR$_1$ phase. Subsequently, a first-order phase transition drives the system into the BR$_{f2}$ phase. Finally, a full $\mathcal{PT}$ transition occurs, bringing the system into the $\mathcal{PT}$-symmetric confined phase. Notably, this $\mathcal{PT}$ transition exhibits features of the YLES, which is indicated by the yellow line. In the following, we examine the phase diagram in Fig. \ref{fig:phase_diagram1} by dividing our analysis into two parts, the real energy spectrum region and the complex energy spectrum region. It is noteworthy that the phase diagram remains invariant under the sign change of $g$, indicating a {$\mathbb{Z}_2$} symmetry between positive and negative values of $g$.

\section{Real-Energy Spectrum Region}
\label{sec:Second-order phase transition}

To investigate the real-energy regime of non-Hermitian quantum systems, it is essential to introduce biorthogonal quantities as appropriate characterizations. For a non-Hermitian Hamiltonian $\hat{H}$, where $\hat{H} \neq \hat{H}^{\dagger}$, the eigenvalue equations of $\hat{H}$ and $\hat{H}^{\dagger}$ are given by~\cite{Sun2021}:
\begin{eqnarray}
    \hat{H} \vert \psi^R_j \rangle = E_j \vert \psi^R_j \rangle , \quad
    \hat{H}^{\dagger} \vert \psi^L_j \rangle = E^*_j \vert \psi^L_j \rangle,
\end{eqnarray}
where $E_j$ ($E_j^*$) and $\vert \psi^R_j \rangle$ ($\vert \psi^L_j \rangle$) are the eigenvalues and the corresponding right (left) eigenvectors of $\hat{H}$, respectively.
The eigenvectors obey the biorthonormal relation and the completeness relation, 
\begin{eqnarray}
    \langle \psi_i^L\vert \psi^R_j \rangle = \delta_{ij}, \quad \sum_j \vert \psi^R_j \rangle \langle \psi_j^L\vert = 1.
\end{eqnarray}
In addition, if the system can be divided into two parts, denoted $A$ and $B$, then the von Neumann entanglement entropy $S$ of the system is defined as $S_A = -{\rm Tr} [\rho_A {\rm ln}\rho_A ]$~\cite{Eisert2010}, where $\rho_A = {\rm Tr}_B(\rho) = {\rm Tr}_B(\vert \psi_j \rangle \langle \psi_j \vert)$.  It is known that the entanglement entropy of a one-dimensional system with length $N$ at a critical point scales as~\cite{Calabrese2004}: {
\begin{eqnarray}
\label{eq:central charge}
    S_A \propto \frac{c_{\rm eff}}{3} {\rm ln} N,
\end{eqnarray}
where $c_{\rm eff}$ is the effective central charge that characterizes the CFT describing the critical point. For the unitary CFT $c_{\rm eff}$ = $c$, while for non-unitary critical point $c_{\rm eff}$ may differ from the central charge $c$ but still governs finite-size and entanglement scaling.}

Within the biorthogonal quantum mechanics framework, the biorthogonal density matrix~\cite{Chang2020,Ye2024,Lu2024} 
\begin{eqnarray}
\label{eq:biorho}
    \rho^{RL} = \frac{\vert \psi^R_j \rangle \langle \psi^L_j \vert} {{\rm Tr}(\vert \psi^R_j \rangle \langle \psi^L_j \vert)}
\end{eqnarray}
more faithfully captures the non-Hermitian characteristics of the system. In order to explore the existence of second-order phase transitions and extract the central charge of the system, we compute the biorthogonal entanglement entropy and analyze its scaling behavior with system size, as shown in Fig. \ref{fig:entropy}. By fixing $g=0.1$, we observe a pronounced peak of the biorthogonal entanglement entropy  in Fig. \ref{fig:entropy}(a) at $m_c=-0.651$ as $m$ is varied. A finite-size scaling analysis at this critical point yields a central charge of $c = 0.483\pm0.019$, consistent with the central charge $c =1/2$ of the Ising universality class.

Next, we adopt a nonequilibrium approach to determine the correlation-length critical exponent. The time-evolved states $\vert \psi_j^R (m_f,m_i,t) \rangle$ and $\vert \psi_j^L (m_f,m_i,t) \rangle$ after a quench from $m_i$ to $m_f$ are obtained as,
\begin{eqnarray}
    \vert \psi_j^R (m_f,m_i,t) \rangle &=& e^{-i\hat{H}(m=m_f)t} \vert \psi_j^R (m_i) \rangle ,\\
    \vert \psi_j^L (m_f,m_i,t) \rangle &=& e^{-i\hat{H}^{\dagger}(m=m_f)t} \vert \psi_j^L (m_i) \rangle,
\end{eqnarray}
by evolving the initial right eigenstates $\vert \psi_j^R (m_i) \rangle$ and left eigenstates $\vert \psi_j^L (m_i) \rangle$ from time $t = 0$. In the following, we will focus on the time evolution of ground states $\vert \psi_0^R (m_i) \rangle$ and $\vert \psi_0^L (m_i) \rangle$ and introduce the
biorthogonal Loschmidt echo as,
\begin{eqnarray}
\label{eq:BLE}
    \mathcal{L} \!\!= \!\!\langle\! \psi_0^L(m_i)\! \vert\! \psi_0^R(m_f,\!m_i,\!t) \!\rangle\! \langle\! \psi_0^L(m_f,\!m_i,\!t) \!\vert \!\psi_0^R(m_i)\! \rangle.
\end{eqnarray}
It has been shown that the decay of the Loschmidt echo can be enhanced by the equilibrium quantum criticality~\cite{Quan2006}. The first minimum of the Loschmidt echo at the time $t_{\rm min,1}$ has recently been shown to scale as~\cite{Hwang2019}
\begin{eqnarray}
\label{eq:LE_scaling}
    1-\mathcal{L}_{\rm min} \propto \delta m^2 N^{2/\nu},
\end{eqnarray}
at the equilibrium critical point for second-order phase transitions. Here $\delta m$ is the small constant step defined by $\delta m = m_f - m_i$. The dynamical scaling law in Eq. (\ref{eq:LE_scaling}) that governs the critically enhanced decay behavior of the Loschmidt echo with respect to $N$ can be used to extract the equilibrium correlation-length critical exponent $\nu$. It has been observed that the minimum of the Loschmidt echo $\mathcal{L}_{\rm min}$, obeys the scaling law described in Eq. (\ref{eq:LE_scaling}) when the system undergoes a small quench near the equilibrium critical point. However, this scaling behavior makes it challenging to employ the Loschmidt echo as a diagnostic tool for equilibrium criticality in the absence of precise knowledge of the critical point. To address this issue, we propose a complementary approach based on the short-time average of the rate function~\cite{Tang2022},
\begin{eqnarray}
\label{eq:ratef}
    \overline{r} = -\frac{1}{N} \frac{{\rm ln}(\overline{\mathcal{L}})}{\delta m^2},
\end{eqnarray}
which is analogous to the ground-state fidelity susceptibility to find $\mathcal{L}_{\rm min}$. Here $\overline{\mathcal{L}}$ is the short-time average of the Loschmidt echo within the time duration $T$ defined by
\begin{eqnarray}
    \overline{\mathcal{L}} = \frac{1}{T} \int_0^T \mathcal{L}dt.
\end{eqnarray}
When investigating second-order phase transitions in physical systems through dynamical approaches, the short-time average rate function serves as a valuable tool for identifying critical transition points. We first calculate the short-time average rate functions $\overline{r}$ from Eq. (\ref{eq:ratef}) by varying the coupling $m$ and find the pseudocritical points $m^{(N)}$, which are derived from the peaks of short-time average rate functions for each lattice $N$ as shown in Fig. \ref{fig:Loschmidt echo}(a). Subsequently, we perform a quench on the system from the pseudocritical point $m^{(N)}$ to $m_f=m^{(N)}+\delta m$. The results of the Loschmidt echoes $\mathcal{L}$ presented in Fig. \ref{fig:Loschmidt echo}(b) exhibit a decay and revival dynamics. The first minima of the Loschmidt echoes $\mathcal{L}_{\rm min}$ are plotted in Fig. \ref{fig:Loschmidt echo}(c) with respect to the lattice size $N$. According to the scaling law in Eq. (\ref{eq:LE_scaling}), we obtain the critical exponent $\nu=1.028\pm0.031$ for $g=0.1$. This result is consistent with the correlation length critical exponent obtained from the equilibrium approach using the biorthogonal fidelity susceptibility in Appendix~\ref{APPENDIX C}.

\begin{figure}[tb]
\centering
\includegraphics[width=0.98\columnwidth]{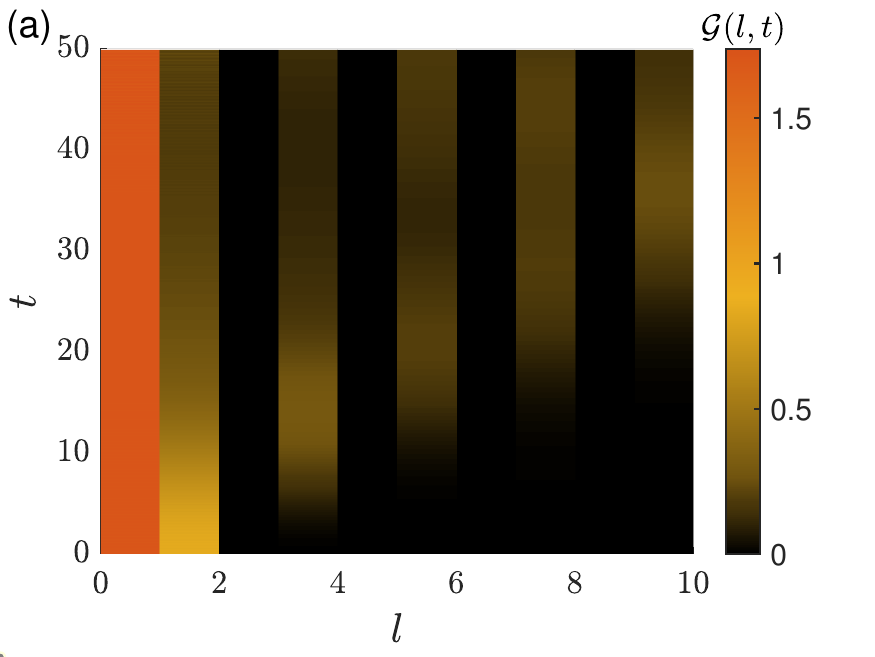}
\includegraphics[width=0.98\columnwidth]{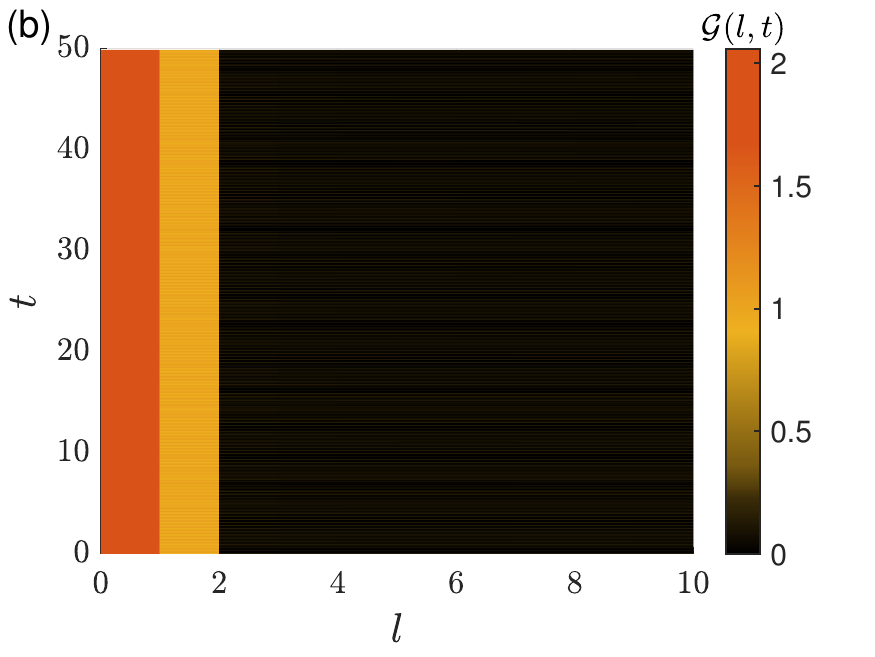}
 \caption{The time evolution of $\mathcal{G}(l,t)$ as a function of $l$ and $t$ for $N = 20$: (a) $m = -5$, $g = 0.1$; (b) $m = 5$, $g = 0.1$.}
 \label{fig:correlation function G}
\end{figure} 

\begin{figure*}[tb]
\centering
\includegraphics[width=0.66\columnwidth]{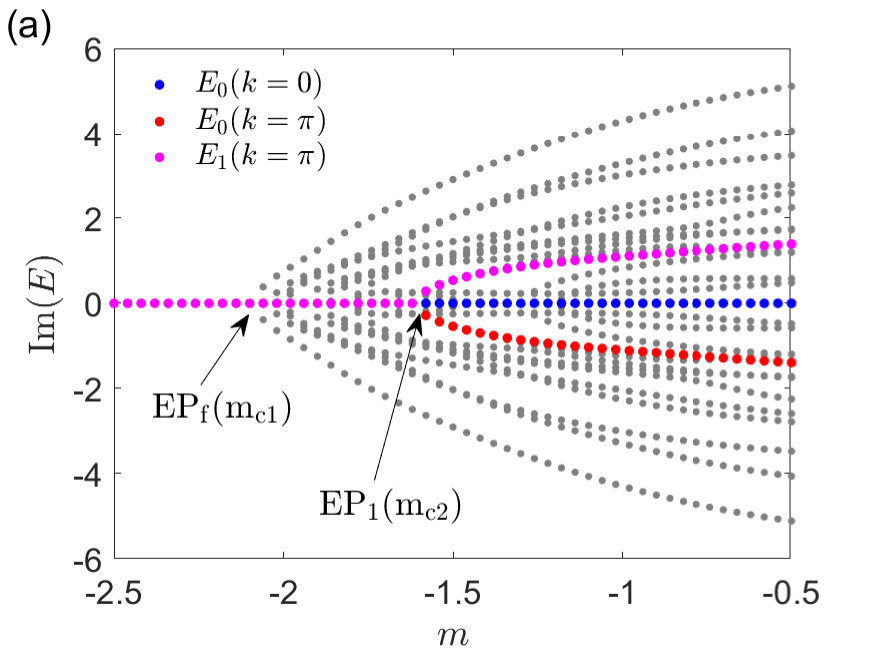}
\includegraphics[width=0.66\columnwidth]{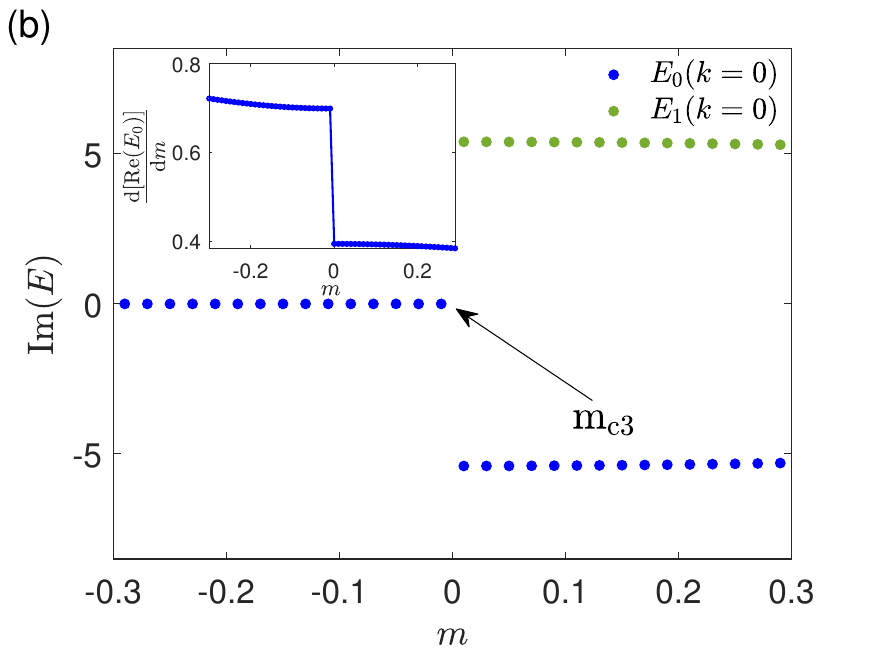}
\includegraphics[width=0.66\columnwidth]{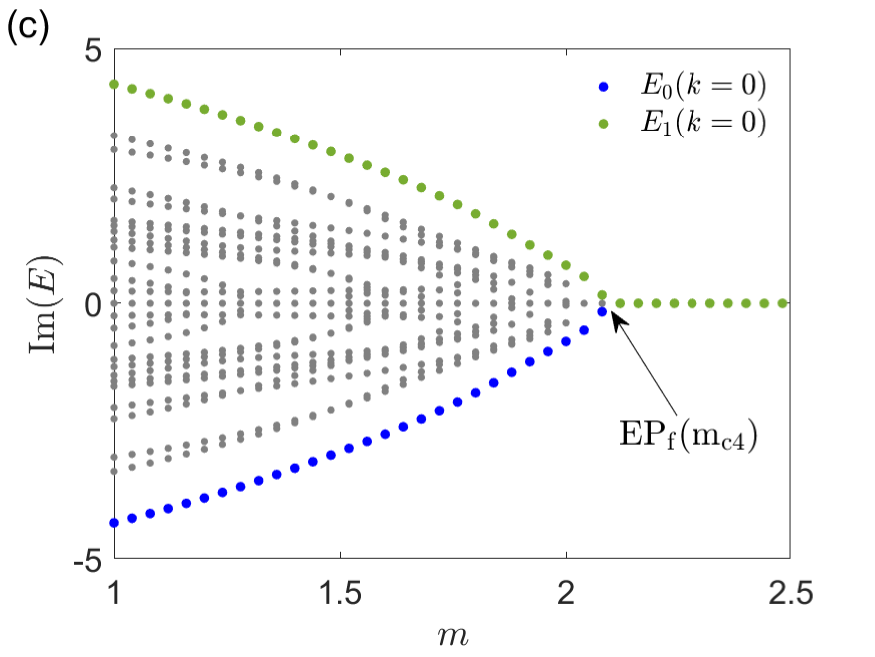}
 \caption{ Energy spectrum as a function of $m$ for $g=1.5$. (a) Imaginary parts of the energy spectrum. The blue circles denote the ground state in the subspace of momentum $k=0$, the red circles represent the ground state for $k=\pi$, and the pink circles indicate the first excited state for $k=\pi$. Here, EP$_{\rm f}$ denotes the EP associated with the full $\mathcal{PT}$ transition, corresponding to the parameter $m_{c1}$, while EP$_{\rm 1}$ refers to the EP characterizing the first excited-state $\mathcal{PT}$ transition, corresponding to $m_{c2}$. (b) Imaginary parts of the energy spectrum. The blue circles denote the ground state for $k=0$, and the green circles represent the first excited state for $k=0$. The point $m_{c3}$ indicates a first-order phase transition. Inset shows the first derivative of the real part of the ground-state energy as a function of $m$. The point of discontinuity in the first derivative corresponds to $m_{c3}$ in panel (b). (c) Imaginary parts of the energy spectrum. The blue circles denote the ground state for $k=0$, and the green circles represent the first excited state for $k=0$. Here, EP$_{\rm f}$ denotes the EP associated with the full $\mathcal{PT}$ transition, corresponding to the parameter $m_{c4}$. Notably, this EP also marks the YLES.}  
 \label{fig:complex_energy}
\end{figure*} 

Moreover, the energy gap at the phase transition point exhibits the following scaling behavior,
\begin{eqnarray}
    \Delta_E \sim N^{-z},
\end{eqnarray}
where $\Delta_E$ is the energy gap between the ground state and the first excited state and $z$ is the dynamical critical exponent. Figure \ref{fig:Loschmidt echo}(d) shows that fitting $\Delta_E$ with respect to the system size $N$ yields $z = 0.986\pm0.034$. The critical exponents $\nu = 1$ and $z = 1$ derived from the finite-size scaling indicate the phase transition for $g = 0.1$ is a second-order phase transition with the Ising universality.

We then proceed to characterize the phases on either side of
the second-order phase transition line. We define a correlation
function
\begin{eqnarray}
    \mathcal{G}(l,t)\! = \!\sum_j\! \sqrt{\langle \psi^L \vert \hat{A}_j \hat{A}_{j+l} \vert \psi^R(t) \rangle \langle \psi^L(t) \vert \hat{A}_j \hat{A}_{j+l} \vert \psi^R \rangle},  \quad
\end{eqnarray}
where $\hat{A}_j = \hat{n}_j - \overline{n}_j$. $\overline{n}_j$ denotes the mean density at site $j$ in the ground state. $\mathcal{G}(l,t)$ is always a localized function of $l$ in the confined phase and is an extended function in the deconfined phase. Figure \ref{fig:correlation function G} illustrates the dynamical behavior of the correlation function $\mathcal{G}(l,t)$ after a local spin excitation is introduced into the ground state of the Hamiltonian (\ref{eq:mathcal_H}). When the parameter $m$ takes
positive values, the correlation remains confined to a limited
spatial region throughout the evolution, reflecting a characteristic feature of the confined phase. Conversely, as $m$ decreases and enters the deconfined regime, $\mathcal{G}(l,t)$ exhibits rapid expansion, spreading across the entire system within a short time scale. {Note that the numerical methods discussed in this section are valid only in the real-spectrum regime.}

\section{Complex-Energy Spectrum Region and Yang-Lee edge singularity}
\label{sec:PT transition}

{In non-Hermitian systems, the presence of $\mathcal{PT}$ symmetry does not guarantee a completely real energy spectrum, as $\mathcal{PT}$ symmetry can undergo breaking.} As a result, a transition that breaks $\mathcal{PT}$ symmetry may occur at an EP, marking a boundary between a $\mathcal{PT}$ symmetric phase with entirely real eigenvalues and a $\mathcal{PT}$-symmetry-broken phase where complex eigenvalues emerge. When this transition involves the entire many-body spectrum, it is referred to as a full $\mathcal{PT}$ transition. Alternatively, if the ground state remains real while the imaginary components first appear in the low-lying excited states, such as the first and second excited levels, this signals a first-excited-state $\mathcal{PT}$ transition.

\begin{figure*}[tb]
\centering
\includegraphics[width=0.66\columnwidth]{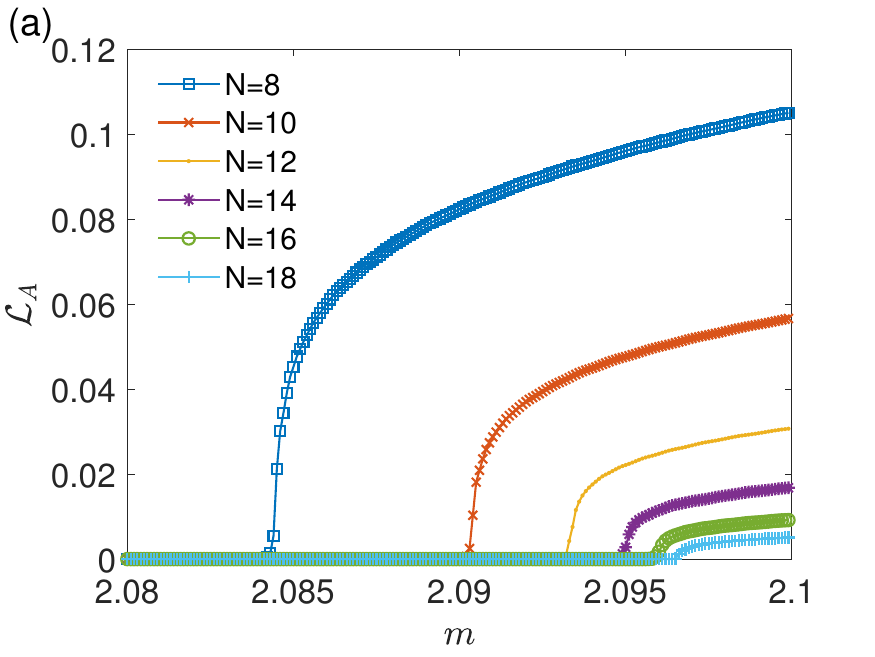}
\includegraphics[width=0.66\columnwidth]{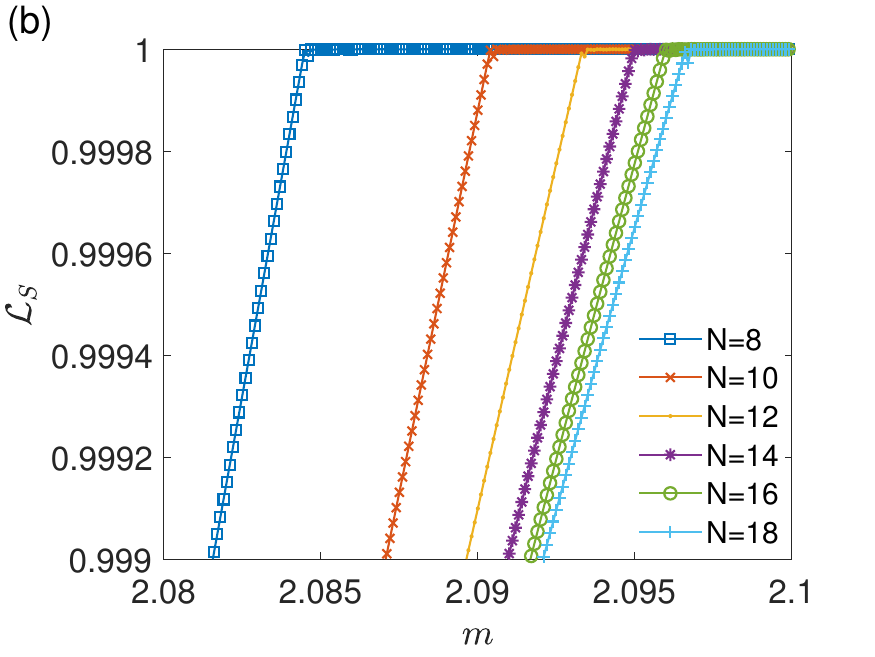}
\includegraphics[width=0.66\columnwidth]{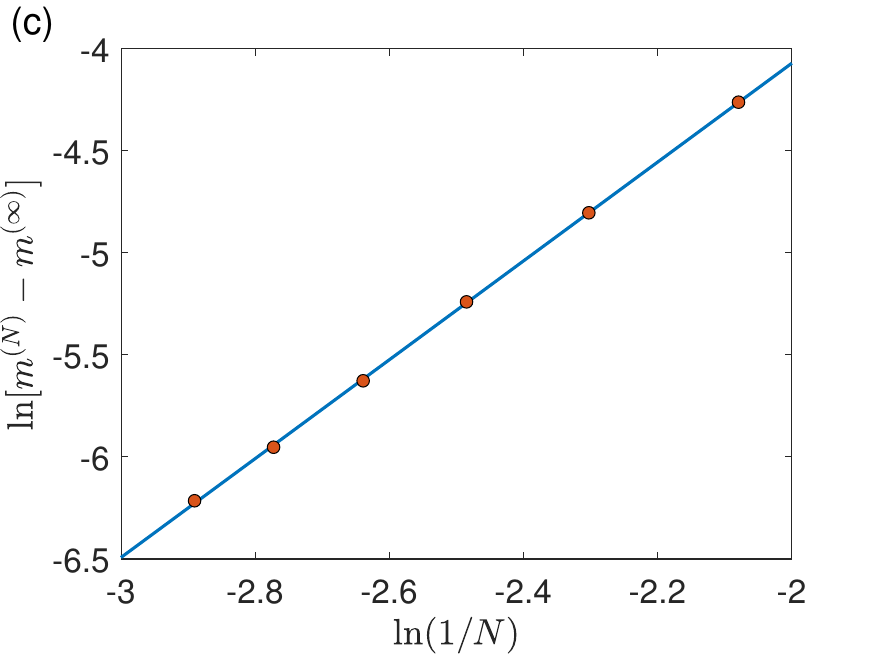}
 \caption{ Finite-size scaling of the critical exponent $\beta$. (a), (b) The associated-biorthogonal Loschmidt echo (\ref{eq:LA}) and the self-normal Loschmidt echo (\ref{eq:SLE}) as a function of $m$. Here, $t=150$, $\delta t=1$ and $\delta m=0.0001$ are used. We select the state with the lowest real part of the energy and a negative imaginary part as the ground state in the complex-energy regime. (c) Finite-size scaling analysis of the difference between the pseudocritical point $m^{(N)}$ and the critical point in the thermodynamic limit $m^{(\infty)}$ as a function of the inverse system size $1/N$ at $g=1.5$. The circles indicate the extracted critical points $m^{(N)}$ for different system sizes, while the solid line corresponds to the fitting curve obtained using Eq. (\ref{eq:Yang-Lee critical exponent}) with an exponent $\beta=2.422\pm0.039$.}
 \label{fig:YL_scaling}
\end{figure*}

To further investigate the $\mathcal{PT}$-symmetry breaking behavior in both the full many-body spectrum and the first excited state, we evaluate the imaginary components of the energy spectrum in the subspace of momentum $k=0$ and $k=\pi$ at $g=1.5$. As shown in Fig. \ref{fig:complex_energy}(a), we find that all energy levels are real when $m<m_{c1}$, while complex energy levels appear when $m_{c1}<m<0$. For $m<m_{c1}$, the system preserves $\mathcal{PT}$ symmetry, as evidenced by an entirely real energy spectrum. When $m$ increases to $m_{c1}$, complex eigenvalues start to emerge, indicating a global $\mathcal{PT}$ transition. However, at this stage, the real parts of the lowest three energy levels remain purely real. As $m$ further increases beyond $m_{c2}$, the first and second excited states (ground states that are degenerate in their real parts at momentum $k=\pi$) separate from the ground state and acquire symmetric imaginary parts, signaling the onset of a first-excited-state $\mathcal{PT}$ transition. Figure \ref{fig:complex_energy}(b) shows a first-order phase transition at $m=m_{c3}=0$, characterized by a discontinuity in the first derivative of the real part of the ground-state energy. Interestingly, the energy levels at $m= m_{c3}$ are purely imaginary. More generally, along the line $m=0$ with $g>1$, all eigenvalues become purely imaginary. At the special point $m=0$ and $g=1$, 
the entire spectrum collapses to zero, marking a multicritical point with complete spectral coalescence.
 
As $m$ increases further, when $m>m_{c4}$, the imaginary parts of the energies vanish, as shown in Fig. \ref{fig:complex_energy}(c), indicating that a full $\mathcal{PT}$ transition occurs at $m=m_{c4}$. In the region $0<m<m_{c4}$, the energies of the ground state and the first excited state are degenerate in their real parts and form complex conjugate pairs in their imaginary parts, while for $m>m_{c4}$, the spectrum becomes purely real. These features indicate clear signatures of the YLES. {This transition is accompanied by a change in the system's dynamical behavior, analogous to the behavior observed in nonintegrable systems analyzed by non-unitary CFT~\cite{Sanno2022,Gao2024,lu2025}.} 

We initially intended to characterize the phase transition using the biorthogonal Loschmidt echo (\ref{eq:BLE}), a natural candidate due to the biorthogonal structure of non-Hermitian systems. While this method successfully captures transitions when the energy spectrum is entirely real, it breaks down in the complex-energy regime. The core issue lies in the definition of the biorthogonal Loschmidt echo, which enforces unitary-like dynamics inconsistent with the intrinsically non-unitary evolution of systems governed by complex spectra. As a result, it fails to provide meaningful physical insights in this regime. To describe the time evolution of the Loschmidt echo in the complex-energy regime within the biorthogonal framework, we introduce a set of associated states~\cite{Jing2024,lu2025}. For a given right eigenstate $\vert \psi_0^R \rangle$, it is expressed as a linear combination of multiple eigenstates. The corresponding left-associated state $\vert \tilde{\psi}_0^L \rangle$ is then defined using the operation
\begin{eqnarray}
    \vert \psi_0^R \rangle = \sum_n c_n \vert \phi_n^R \rangle \rightarrow  \vert \tilde{\psi}_0^L \rangle = \sum_n c_n \vert \phi_n^L \rangle,
\end{eqnarray}
where the coefficient $c_n=\langle \phi_n^L \vert \psi_0^R \rangle$ is the same as in $\langle \tilde{\psi}_0^L \vert = \Sigma_n c_n^* \langle \phi_n^L \vert$. Here, $\vert \phi_n^R \rangle$ and $\vert \phi_n^L \rangle$ form a biorthogonal basis and $\vert \tilde{\psi}^L \rangle$ denotes the associated state. The biorthogonal Loschmidt echo in the associated state bases, referred to as the associated-biorthogonal Loschmidt echo, is defined as
\begin{eqnarray}
    \mathcal{L}_A\!=\!\langle \! \tilde{\psi}_0^L \! (\!m_i\!) \vert \psi_0^R \!(\!m_f\!,\!m_i\!,\!t\!) \rangle \langle \tilde{\psi}_0^L\! (\!m_f\!,\!m_i\!,\!t\!) \vert \psi_0^R \!(\!m_i\!) \rangle.
    \label{eq:LA}
\end{eqnarray}
To make the location of the phase transition more evident, we present the logarithm of the associated-biorthogonal Loschmidt echo, as shown in Fig. \ref{fig:YL_scaling}(a). A clear change in $\mathcal{L}_A$ is observed at the transition point, which further supports that the associated-biorthogonal Loschmidt echo effectively captures phase transitions in the complex-energy regime.

In fact, the {self-normal} Loschmidt echo can also capture the YLES~\cite{lu2025} and is experimentally accessible~\cite{Jeanneret2014,Swingle2016,anchez2020}. One can define the self-normal Loschmidt echo:
\begin{eqnarray}
\label{eq:SLE}
    \mathcal{L}_S \!\!=\!\! \langle\! \psi_0^R(m_i)\! \vert\! \psi_0^R(m_f,\!m_i,\!t) \!\rangle\! \langle\! \psi_0^R(m_f,\!m_i,\!t) \!\vert \!\psi_0^R(m_i)\! \rangle.
\end{eqnarray}
The results of the self-normal Loschmidt echo $\mathcal{L}_S$ are presented in Fig. \ref{fig:YL_scaling}(b). A pronounced decay of the self-normal Loschmidt echo is observed at the $\mathcal{PT}$-symmetry-broken phase.

To further validate that the transition corresponds to the YLES, we analyze the finite-size scaling at its boundary.  Near the YLES, the pseudocritical point $m^{(N)}$ approaches its thermodynamic value $m^{(\infty)}$ according to the universal scaling form
\begin{eqnarray}
    \label{eq:Yang-Lee critical exponent}
    m^{(N)} = m^{(\infty)} + aN^{-\beta},
\end{eqnarray}
where  $\beta$  is the {magnetic scaling exponent}, taking the exact value $\beta = 12/5$ according to non-unitary CFT~\cite{Uzelac1979,Gehlen_1991,Cardy1985}. {This exponent governs the finite-size shift of the pseudocritical point and, equivalently, determines the universal power-law approach of the leading Yang-Lee zeros toward the real axis as the thermodynamic limit is approached~\cite{Michael1978,Cardy2023}.} Figure \ref{fig:YL_scaling}(c) shows our finite-size scaling results, where an excellent data collapse is obtained with a fitted value $\beta=2.422\pm0.039$, in very good agreement with the exact prediction. {This numerical result unambiguously identifies the EP $m_{c4}$ as the Yang-Lee edge and confirms that the transition is a second-order critical point~\cite{Kim2006,Wei2017,Yin2017}.}

{To corroborate the CFT description of the YLES, we also extract the effective central charge at the edge. At the YLES, the low-energy theory is the non-unitary CFT with central charge $c=-22/5$,  for which the effective central charge takes the exact value $c_{\rm eff}=2/5$~\cite{Sanno2022}. Approaching the edge from the $\mathcal{PT}$-symmetric side, we fit the biorthogonal entanglement entropy using Eq.~(\ref{eq:central charge}) and obtain $c_{\rm eff} = 0.4083 \pm 0.0785$, which agrees with the exact YLES prediction $c_{\rm eff}=2/5$ within numerical uncertainty. These results provide direct evidence that the transition belongs to the YLES universality class. See Appendix~\ref{APPENDIX D} for details of the finite-size scaling analysis.}

\begin{figure}[tb]
\centering
\includegraphics[width=\columnwidth]{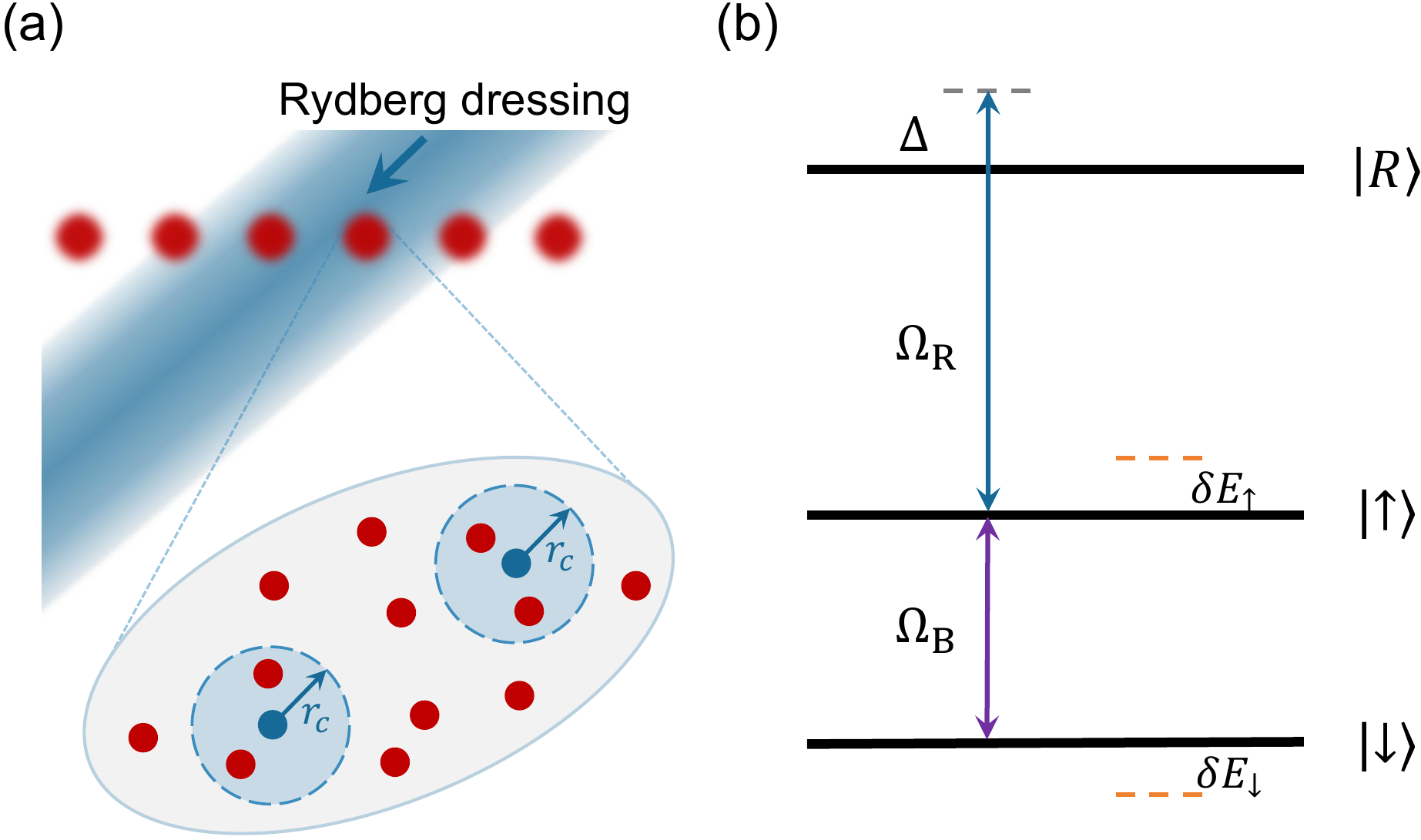}
\caption{Proposed experimental scheme. (a) One-dimensional array of optical microtraps loaded with cesium atoms; Rydberg dressing with 319\,nm light generates effective Ising interactions (a characteristic length scale $r_c$ is indicated for reference). (b) {Schematic level diagram showing the two hyperfine ground states that encode a qubit and their off-resonant coupling to a Rydberg state with single-photon Rabi frequency $\Omega_R$ and detuning $\Delta$; a Raman (or microwave) drive with Rabi frequency $\Omega_B$ implements the transverse field, while differential light shifts set the longitudinal field $h_z$, together enabling tunable ferromagnetic Ising interactions.}}
\label{fig:Experimental proposal}
\end{figure}

We next demonstrate that the target non-Hermitian Hamiltonian [Eq. (\ref{eq:Ising})] could be experimentally realized, allowing for the observation of the YLES in Rydberg-atom arrays. We show that Eq. (\ref{eq:Ising}), through a similarity transformation (\ref{eq:transformation}), can be transformed into the following form:
\begin{eqnarray}
  \label{eq:tranformedIsing}
    \tilde{H} \!= \!J \!\sum_{j=1}^{N} \!\tilde{\sigma}^z_j\tilde{\sigma}^z_{j+1} \!+\!\sqrt{1-g^2} \!\sum_{j=1}^N \!\tilde{\sigma}^x_j\! +\!h_z \!\sum_{j=1}^{N} \! \tilde{\sigma}^z_j, 
\end{eqnarray}
{which is Hermitian for $\vert g \vert \le 1$ but becomes genuinely non-Hermitian for $\vert g \vert > 1$, where the YLES is located.  
All terms in Eq. (\ref{eq:tranformedIsing}) are implementable with established techniques,  as illustrated in Fig. \ref{fig:Experimental proposal}.}

Following Refs.~\cite{Borish2020,Ren2022}, we propose a scheme based on a one-dimensional array of optical microtraps loaded with cesium atoms [Fig.~\ref{fig:Experimental proposal}(a)].  The qubit is encoded in the hyperfine ground states $\lvert \downarrow \rangle = \lvert 6S_{1/2}, F=3, m_F=+3\rangle$ and $\lvert \uparrow \rangle = \lvert 6S_{1/2}, F=4, m_F=+4\rangle$. As shown in Fig.~\ref{fig:Experimental proposal}(b), a differential light shift induced by off-resonant beams produces a longitudinal field $h_z = (\delta E_\uparrow - \delta E_\downarrow)/2$. Ferromagnetic Ising interactions $J$ are generated by Rydberg dressing, wherein $\lvert \uparrow \rangle$ is off-resonantly coupled to a Rydberg level $\lvert R \rangle = \lvert 60P_{3/2}, m_J = +3/2\rangle$ with single-photon Rabi frequency $\Omega_R$ and detuning $\Delta$~\cite{Shen2023}. These dressing parameters set a characteristic interaction length scale  $r_c$ and $\Delta>0$ yields ferromagnetic interactions. The transverse field $\propto \tilde{\sigma}^x$  is implemented by a resonant Raman (or microwave) drive between $\lvert \uparrow\rangle$ and $\lvert \downarrow\rangle$ with tunable Rabi frequency $\Omega_B$~\cite{Bernien2017}.

To access the non-Hermitian regime $\vert g\vert >1$, an effective imaginary transverse coupling can be engineered using a weak measurement combined with post-selection on null outcomes. Concretely, by continuously monitoring the projector $(1+\tilde{\sigma}^{x}_{j})/2$ at rate $\kappa$ and selecting trajectories with no detected quantum jumps, the stochastic evolution reduces to an effective non-Hermitian term $- i \frac{\kappa}{4} \tilde{\sigma}^{x}_{j}$ in the Hamiltonian (up to an irrelevant constant shift), i.e., an imaginary transverse field whose strength $g$ is set by the measurement rate. Here $g$ can be interpreted either as a dissipation rate or as a continuous measurement strength that tunes the system across the $\mathcal{PT}$-symmetric and $\mathcal{PT}$-broken regimes~\cite{PhysRevB.109.104312,Daley2014}. Such a scheme can be implemented by optically pumping one of the $\tilde{\sigma}^{x}$ eigenstates to a fast-decaying auxiliary level while monitoring emitted photons, or equivalently by engineering a heralded loss channel. Tuning $\kappa$ allows $g$ to cross the EP, thereby driving the system into the  $\mathcal{PT}$-broken phase where the YLES occurs.
Once the Hamiltonian is realized, the critical exponent can be extracted by measuring the self-normalized Loschmidt echo [Eq. (\ref{eq:SLE})], which has already been demonstrated to be experimentally accessible in Rydberg platforms.

\section{Summary and conclusions}
\label{sec:Summary and conclusions}

In this work, we explored quantum criticality in a non-Hermitian detuned PXP model and established a comprehensive phase diagram. Starting from a numerically identified phase transition point, we construct an exact second-order phase transition boundary through a similarity transformation. To characterize this phase transition, we employed a combination of biorthogonal entanglement entropy and the biorthogonal Loschmidt echo, confirming from both static and dynamic perspectives that the critical behavior belongs to the Ising universality class. {These two biorthogonal observables are well-defined only in the real-spectrum regime.} In the $\mathcal{PT}$-symmetric region, we used the correlation function to reveal a confinement–deconfinement transition, identifying distinct dynamical features in each phase. Moving into the complex-energy regime, we uncovered two distinct types of $\mathcal{PT}$ symmetry breaking, a full spectrum transition and a transition involving the first excited state. Furthermore, by introducing the associated-biorthogonal Loschmidt echo and the self-normal Loschmidt echo, we located the YLES and extracted the corresponding critical exponent, in agreement with the theoretical predictions from non-unitary CFT. Finally, we propose an experimental scheme to observe the YLES in Rydberg atomic arrays, providing a feasible pathway for exploring non-Hermitian critical behavior and singularities in future experiments.

\begin{acknowledgments}
The authors appreciate very insightful discussions with M. C. Lu, C. Z. Lu and J. C. Tang. This work is supported by the National Natural Science Foundation of China (NSFC) under Grant No. 12174194, the Fundamental Research Funds for the Central Universities under Grant No.
NS2023055, the NSFC under Grant No. 11704186, Postgraduate Research \&
Practice Innovation Program of Jiangsu Province, under Grant No. KYCX23\_0347 and No. KYCX25\_0640, Opening Fund of the Key Laboratory of Aerospace Information Materials and Physics (Nanjing University of Aeronautics and Astronautics), MIIT, Top-notch Academic Programs Project of Jiangsu Higher Education Institutions (TAPP), and stable supports for basic institute research under Grant No. 190101. 
\end{acknowledgments}

\section*{Data availability}
The data that support the findings of this article are openly available~\cite{data}.

\appendix

\section{Non-Hermitian Ising chain with transverse and longitudinal fields}
\label{APPENDIX A}
We start from the non-Hermitian Ising chain with transverse and longitudinal fields,
\begin{equation}
    \hat{H}
    = h_x \sum_{j=1}^N \sigma_j^x
    + g e^{i\alpha} \sum_{j=1}^N \sigma_j^y
    + J \sum_{j=1}^N \sigma_j^z \sigma_{j+1}^z
    + h_z \sum_{j=1}^N \sigma_j^z .
    \label{eq:appA_Ising}
\end{equation}
For convenience we introduce
\begin{equation}
    n_j \equiv \frac{1+\sigma_j^z}{2}, \qquad
    P_j \equiv \frac{1-\sigma_j^z}{2},
\end{equation}
so that $n_j = 1 - P_j$.  
Writing $h_z-2J\equiv 2m$, one may rewrite the Ising and longitudinal-field parts in terms of $n_j$ as
\begin{eqnarray}
     &  & J \sum_{j=1}^{N} \sigma_j^z \sigma_{j+1}^z + h_z \sum_{j=1}^{N} \sigma_j^z \nonumber \\
    &= & 4J \sum_{j=1}^{N} n_j n_{j+1}
       + 4m \sum_{j=1}^{N} n_j
       + \mathrm{const.},
    \label{eq:appA_Jhz_identity}
\end{eqnarray}
where the additive constant depends on boundary conditions and is physically irrelevant. Using Eq.~\eqref{eq:appA_Jhz_identity}, Eq.~\eqref{eq:appA_Ising} becomes
\begin{equation}
    \hat{H}
    = h_x \sum_{j} \sigma_j^x
    + g e^{i\alpha} \sum_{j} \sigma_j^y
    + 4m \sum_{j} n_j
    + 4J \sum_{j} n_j n_{j+1}.
    \label{eq:appA_split}
\end{equation}

In the parameter regime where $J$ is the dominant energy scale and the transverse field  $h_x$, non-Hermitian coupling $ g$, and detuning $2m$ are all small and of comparable magnitude—i.e.,  $h_x, g,  \vert 2m \vert \ll J$ with $ h_x \sim g \sim \vert 2m \vert$, we separate $\hat{H}=H_0+V$ with the high-energy part
\begin{equation}
    H_0 = 4J \sum_{j=1}^{N} n_j n_{j+1},
\end{equation}
and treat the remainder $V$ perturbatively. In the interaction picture,
\begin{equation}
    U(t) = e^{-i H_0 t}, \qquad
    O_{\mathrm I}(t) = U^\dagger(t)\, O \,U(t),
\end{equation}
a Baker-Campbell-Hausdorff expansion yields the closed forms
\begin{eqnarray}
    &&U^\dagger \sigma_j^x U \nonumber \\
    &=& \cos\!\big[(n_{j-1}+n_{j+1})(4Jt)\big]\sigma_j^x
     + \sin\!\big[(n_{j-1}+n_{j+1})(4Jt)\big]\sigma_j^y, \nonumber
    \label{eq:appA_sigmax_rot}  
    \end{eqnarray}
    \begin{eqnarray}
    &&U^\dagger \sigma_j^y U \nonumber \\
    &=& - \sin\!\big[(n_{j-1}+n_{j+1})(4Jt)\big] \sigma_j^x
     + \cos\!\big[(n_{j-1}+n_{j+1})(4Jt)\big]\sigma_j^y,\nonumber
    \label{eq:appA_sigmay_rot}
\end{eqnarray}
equivalently
\begin{equation}
    U^\dagger \sigma_j^\pm U
    = e^{\pm i (4Jt)(n_{j-1}+n_{j+1})}\,\sigma_j^\pm,\end{equation}
where  $\sigma_j^\pm \equiv \frac{1}{2}(\sigma_j^x \pm i \sigma_j^y)$ .
Because $h_x \ll J$, rapidly oscillating terms at frequencies $\sim 4J$ and $8J$ average out under the rotating-wave approximation (RWA). Using
\begin{eqnarray}
    &&e^{\pm i (4Jt)(n_{j-1}+n_{j+1})} \nonumber \\
   &=& P_{j-1} P_{j+1}
      + \big(n_{j-1} P_{j+1} + P_{j-1} n_{j+1}\big) e^{\pm i 4Jt} \nonumber \\
   &+& n_{j-1} n_{j+1} e^{\pm i 8Jt},  
\end{eqnarray}
the fast-oscillating contributions are dropped, leaving the slowly varying projections
\begin{equation}
    U^\dagger \sigma_j^x U \approx P_{j-1}\,\sigma_j^x\,P_{j+1},
    U^\dagger \sigma_j^y U \approx P_{j-1}\,\sigma_j^y\,P_{j+1}.
\end{equation}
Therefore, in the interaction picture and within the RWA,
\begin{eqnarray}
    H_{\mathrm I}^{(\mathrm{RWA})}
    &\approx&
    h_x \sum_{j=1}^{N} P_{j-1} \sigma_j^x P_{j+1}
    + g e^{i\alpha} \sum_{j=1}^{N} P_{j-1} \sigma_j^y P_{j+1}
\nonumber \\
&+& 4m \sum_{j=1}^{N} n_j.
    \label{eq:appA_HI_RWA}
\end{eqnarray}

Projecting onto the blockade subspace (no adjacent excitations) replaces the diagonal term by its constrained form
\begin{equation}
    4m \sum_{j=1}^{N} n_j
    \;\longrightarrow\;
    2m \sum_{j=1}^{N} P_{j-1}\, n_j \,P_{j+1} \;+\; \mathrm{const.}
\end{equation}
Dropping the additive constant, we arrive at the constrained Hamiltonian
\begin{eqnarray}
\label{equ:nHdPXP_appA}
    \hat{H}
   && \approx
  \sum_{j=1}^{N}  P_{j-1} \sigma^x_{j} P_{j+1} +ge^{i\alpha} \sum_{j=1}^{N}  P_{j-1}\sigma^y_{j} P_{j+1} \nonumber \\ &&+ 2m \sum_{j=1}^N P_{j-1}n_{j}P_{j+1}.
\end{eqnarray}
which coincides with Eq.~(\ref{equ:nHdPXP}) in the main text. In particular, setting $\alpha=\pi/2$ gives $g e^{i\alpha}= i g$, recovering the non-Hermitian detuned PXP model used throughout.
 
 \section{Phase-diagram slices of the detuned PXP model}
\label{APPENDIX B}
\renewcommand{\thesection}{B}
\setcounter{figure}{0}
\counterwithin{figure}{section}

Then we consider the detuned PXP model [Eq. (\ref{equ:nHdPXP_appA})] parameterized by the phase $\alpha$ in the coefficient of $\sigma^y$, 
\begin{eqnarray}
\label{equ:HdPXP}
\hat{H}
=&& \sum_{j=1}^{N} P_{j-1}\,\sigma^x_{j}\,P_{j+1}
\;+\; g e^{i\alpha} \sum_{j=1}^{N} P_{j-1}\,\sigma^y_{j}\,P_{j+1} \nonumber \\
&&+\; 2m \sum_{j=1}^N P_{j-1}\,n_{j}\,P_{j+1}.
\end{eqnarray}
 The model is Hermitian for $\alpha=0$, and non-Hermitian for generic $\alpha\neq 0,\pi$. The main text focuses on the $\alpha=\pi/2$ slice ($g e^{i\alpha}= i g$).

\begin{figure}[tb]
\centering
\includegraphics[width=\columnwidth]{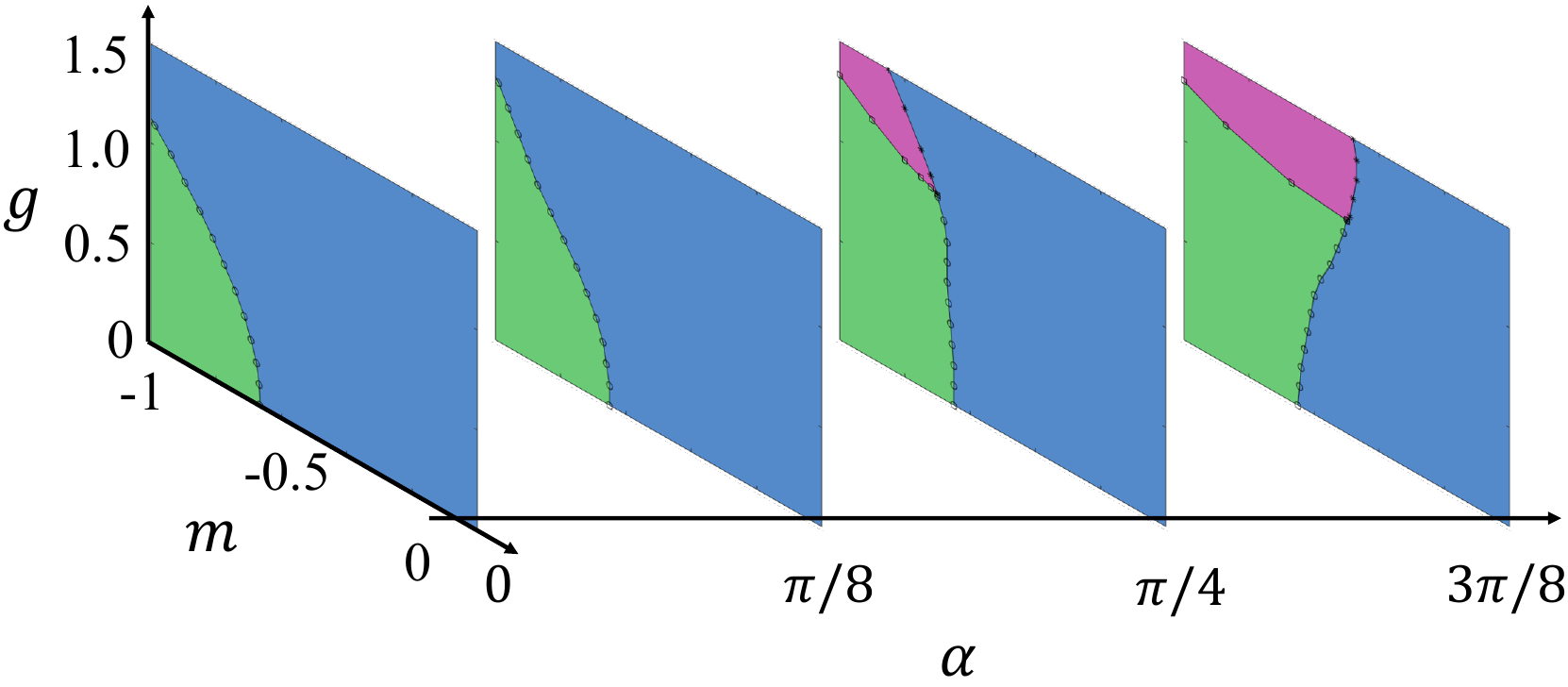}
\caption{Slices of the phase diagram of the detuned PXP model in Eq.~(\ref{equ:HdPXP}) at fixed $\alpha=0,\,\pi/8,\,\pi/4,\,3\pi/8$. Green and blue denote the deconfined and confined phases (ground-state classification), respectively; pink marks the $\mathrm{BR}_1$ regime.}
\label{fig:phase_diagram_alpha_0}
\end{figure}

To illustrate how the phase structure evolves with $\alpha$, Fig.~\ref{fig:phase_diagram_alpha_0} shows phase-diagram slices of Eq.~\eqref{equ:HdPXP} at several fixed $\alpha$. For $\alpha=0$ the Hamiltonian is fully Hermitian and the spectrum is purely real for all real parameters. In this case, the second-order transition line is described analytically by
$m_c = -0.655\sqrt{1+g^2}$.
For intermediate phases $\alpha\in(0,\pi/2)$, the term $g e^{i\alpha}\sum_j P_{j-1}\sigma^y_j P_{j+1}$ contains both Hermitian and anti-Hermitian components; correspondingly, we do not find extended regions where the full many-body spectrum remains purely real for $g\neq 0$.

\begin{figure}[tb]
\centering
\includegraphics[width=0.9\columnwidth]{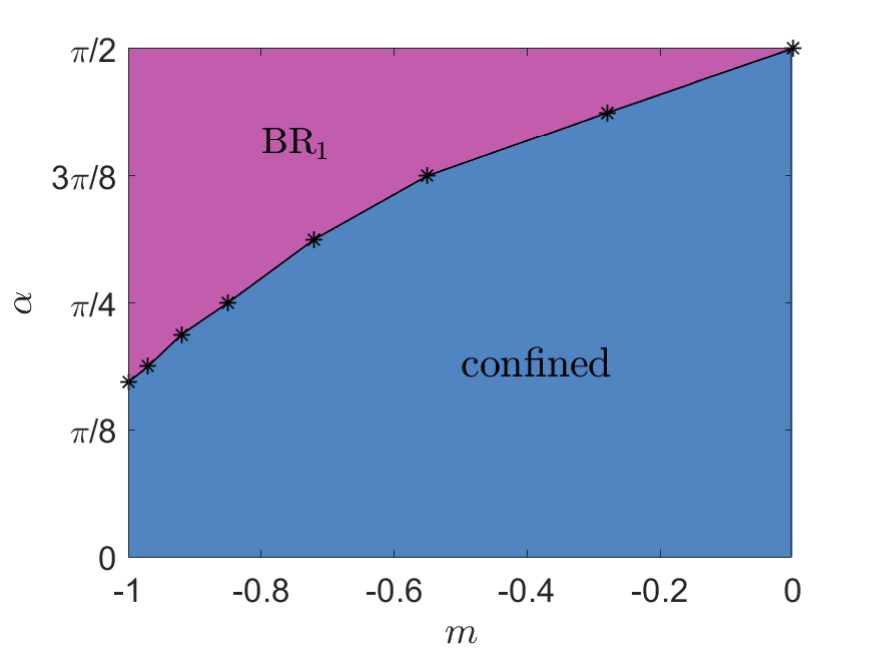}
\caption{{Phase diagram of Eq.~(\ref{equ:HdPXP}) at fixed $g=1.5$. Blue denotes the confined phase; pink marks the $\mathrm{BR}_1$ regime. In this cut a $\mathrm{BR}_1$ region emerges near $\alpha \approx 11\pi/64$.}}
\label{fig:phase_diagram_g_15}
\end{figure}

{Figure~\ref{fig:phase_diagram_g_15} further presents the phase diagram at fixed $g=1.5$ as a function of $\alpha$ and $m$. In this cut, a $\mathrm{BR}_1$ regime (first-excited-state $\mathcal{PT}$ breaking with a real ground state) appears near $\alpha \approx 11\pi/64$.}

\renewcommand{\thesection}{C}
\section{Self-normal and biorthogonal fidelity susceptibility}
\label{APPENDIX C}

\setcounter{figure}{0}
\counterwithin{figure}{section}
\begin{figure}[tb]
\centering
\includegraphics[width=0.87\columnwidth]{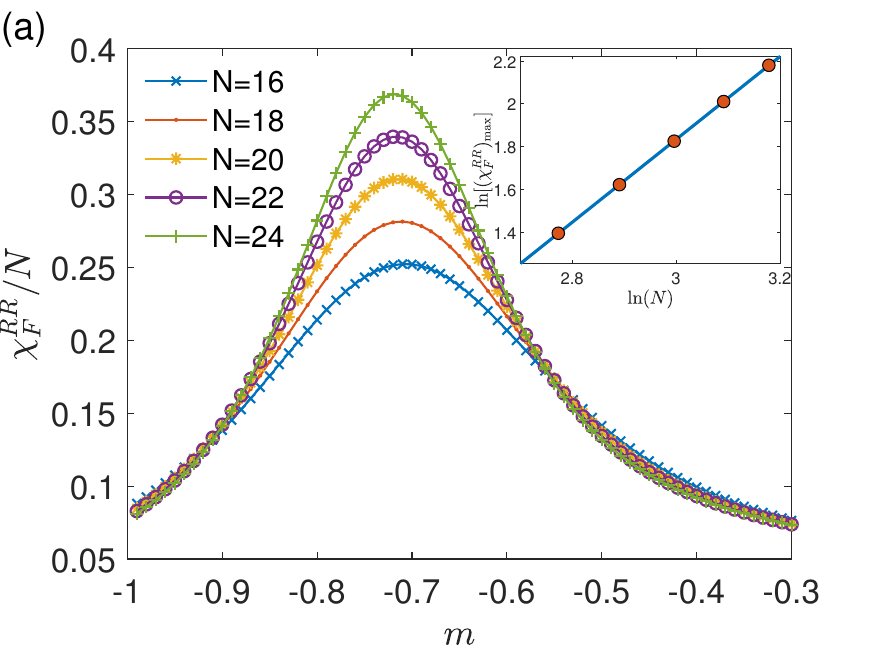}
\includegraphics[width=0.87\columnwidth]{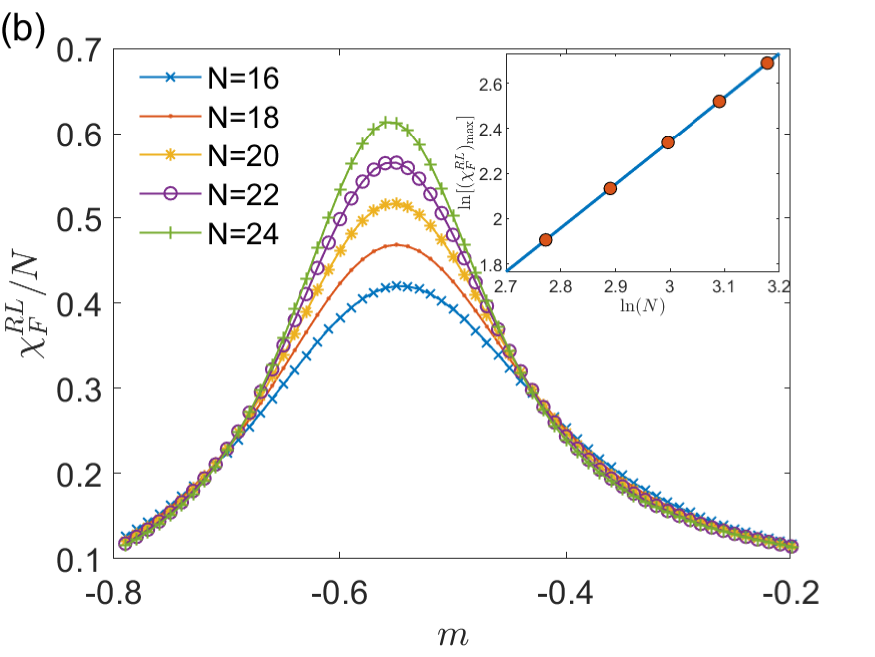}
 \caption{{Determination of correlation length critical exponent by fidelity susceptibility for $g=0.5$. (a) The self-normal fidelity susceptibility $\chi_{F}^{RR}/N$ for Hermitian Hamiltonian (\ref{equ:HdPXP}) with $\alpha=0$. Inset shows the finite-size scaling of the maxima of $\chi_{F}^{RR}$. The correlation length critical exponent obtained from fitting curve is $\nu=1.033\pm0.035$. (b) The biorthogonal fidelity susceptibility $\chi_{F}^{RL}/N$ for non-Hermitian Hamiltonian (\ref{equ:nHdPXP}). Inset shows the finite-size scaling of the maxima of $\chi_{F}^{RL}$. The correlation length critical exponent obtained from fitting curve is $\nu=1.035\pm0.039$.}}
 \label{fig:chiF_scaling} 
\end{figure}

As a function of $m$ across a quantum critical point, the fidelity susceptibility $\chi_F$ develops a broad peak at finite $N$, locating the pseudocritical points $m^{(N)}$. For the extensive fidelity susceptibility, the peak height obeys
\begin{eqnarray}
\chi_F \sim N^{2/\nu},
\end{eqnarray}
so that the \emph{per-site} quantity scales as $\chi_F/N \sim N^{2/\nu-1}$.

Likewise, in non-Hermitian quantum systems, two distinct types of fidelity susceptibility can be defined: the self-normal fidelity susceptibility and the biorthogonal fidelity susceptibility. The Uhlmann fidelity
\begin{eqnarray}
    F = {\rm Tr} \sqrt{\sqrt{\rho(m)}\rho(m+\delta m) \sqrt{\rho(m)}}
\end{eqnarray}
{for the self-normal density matrix and the biorthogonal density matrix can be defined as~\cite{Hauru}:
\begin{eqnarray}
    F^{RR} \!\!&=& \vert \langle \psi^R(m) \vert \psi^R(m+\delta m)\rangle \vert , \\
    F^{RL} \!\!&=& \!\!\sqrt{\!\langle\! \psi^L(m\!+\!\delta m) \!\vert\! \psi^R(m)\! \rangle \!\langle \!\psi^L(m) \!\vert \!\psi^R(m\!+\!\delta m)\! \rangle}.
\end{eqnarray}
The corresponding fidelity susceptibility is then given by~\cite{You2007,Sun2017}:
\begin{eqnarray}
    \chi_{F}^{RR,RL} =   \lim_{\delta m \to 0} \frac{-2 {{\rm ln} F^{RR,RL}}}{\delta m^2}.
\end{eqnarray}
We calculate the self-normal fidelity susceptibility  $\chi_{F}^{RR}$ for Hermitian Hamiltonian (\ref{equ:HdPXP}) with $\alpha=0$ and the biorthogonal fidelity susceptibility $\chi_{F}^{RL}$ for non-Hermitian Hamiltonian (\ref{equ:nHdPXP}), as shown in Fig. \ref{fig:chiF_scaling}. The peaks of the fidelity susceptibility around the critical point become more pronounced for increasing the system sizes $N$. The maxima of fidelity susceptibility show a power-law divergence. Despite differences in the Hamiltonians’ Hermiticity and in the locations of the transitions, we extract the same critical exponent $\nu \approx 1$.}

\renewcommand{\thesection}{D}
\section{Effective central charge}
\label{APPENDIX D}

\setcounter{figure}{0}
\counterwithin{figure}{section}
{We extract the effective central charge $c_{\rm eff}$ at the Yang--Lee edge by finite-size scaling of the biorthogonal entanglement entropy $S_A$ on the $\mathcal{PT}$-symmetric side as the edge is approached. According to Eq.~(\ref{eq:central charge}),
\begin{equation}
S_A(N) \;=\; \frac{c_{\rm eff}}{3}\,\ln N \;+\; \mathrm{const}.
\end{equation}
Figure~\ref{fig:c_eff} shows the fit for Hamiltonian~(\ref{equ:nHdPXP}) at $g=1.5$, from which we obtain
$c_{\rm eff}=0.4083 \pm 0.0785$. For the non-unitary minimal model $\mathcal{M}(2,5)$ describing the Yang--Lee edge singularity, the central charge is $c=-22/5$ and the effective central charge is
$c_{\rm eff}=c-24 h_{\min}=2/5$. Using $c=c_{\rm eff}-24/5$ to convert the fitted value yields
$c=-4.3917 \pm 0.0785$, fully consistent with the CFT prediction.} 
 
\begin{figure}[ht]
\centering
\includegraphics[width=0.87\columnwidth]{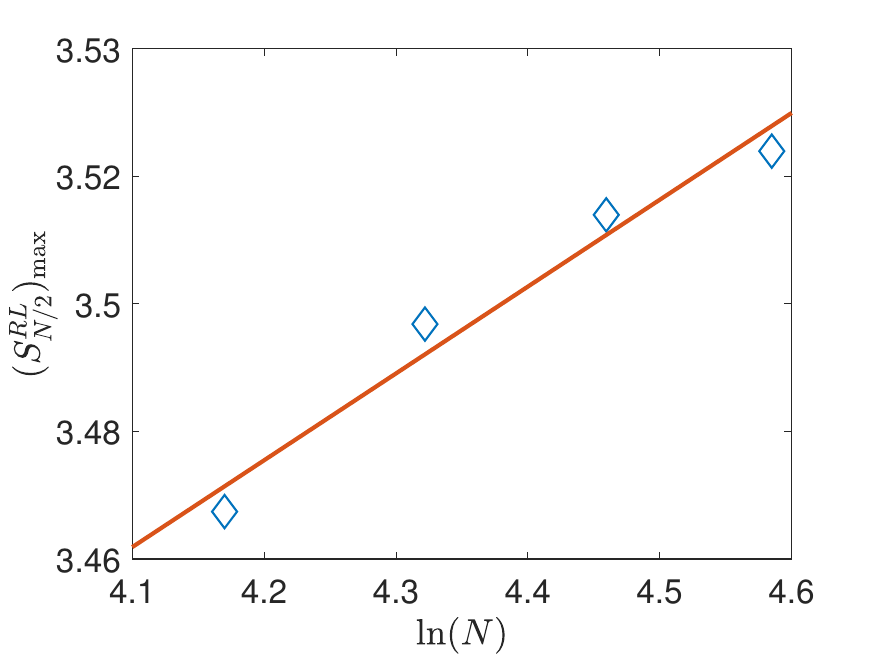}
\caption{Finite-size scaling of the biorthogonal entanglement entropy approaching the YLES from the $\mathcal{PT}$-symmetric side for $g=1.5$ in Hamiltonian~(\ref{equ:nHdPXP}). A fit to Eq.~(\ref{eq:central charge}) gives $c_{\rm eff}=0.4083 \pm 0.0785$, in agreement with the YLES value $c_{\rm eff}=2/5$.}
\label{fig:c_eff}
\end{figure}
\newpage

\bibliography{ref}

\end{document}